\newcommand\apjcls{1}
\newcommand\aastexcls{2}
\newcommand\othercls{3}

\newcommand\papercls{\aastexcls}
\documentclass[tighten, times, twocolumn]{aastex62}  % ApJ look-alike
\if\papercls \apjcls
\usepackage{apjfonts}
\else\if\papercls \othercls
\usepackage{epsfig}
\usepackage{margin}
\usepackage{times}
\fi\fi
\usepackage{ifthen}
\usepackage{natbib}
\usepackage{amssymb, amsmath}
\usepackage{appendix}
\usepackage{etoolbox}
\usepackage[T1]{fontenc}
\usepackage{paralist}

\if\papercls \apjcls
\newcommand\aas{\ref@jnl{AAS Meeting Abstracts}}% *** added by jh
\newcommand\dps{\ref@jnl{AAS/DPS Meeting Abstracts}}% *** added by jh
\newcommand\maps{\ref@jnl{MAPS}}% *** added by jh
\else\if\papercls \othercls
\usepackage{astjnlabbrev-jh}
\fi\fi

\if\papercls \aastexcls
\hypersetup{citecolor=blue, % color for \cite{...} links
            linkcolor=blue, % color for \ref{...} links
            menucolor=blue, % color for Acrobat menu buttons
            urlcolor=blue}  % color for \url{...} links
\else
\usepackage[%pdftex,      % hyper-references for pdflatex
bookmarks=true,           %%% generate bookmarks ...
bookmarksnumbered=true,   %%% ... with numbers
colorlinks=true,          % links are colored
citecolor=blue,           % color for \cite{...} links
linkcolor=blue,           % color for \ref{...} links
menucolor=blue,           % color for Acrobat menu buttons
urlcolor=blue,            % color for \url{...} links
linkbordercolor={0 0 1},  %%% blue frames around links
pdfborder={0 0 1},
frenchlinks=true]{hyperref}
\fi

\if\papercls \othercls

\else

\fi

\providecommand{\adsurl}[1]{\href{#1}{ADS}}

\makeatletter
\patchcmd{\NAT@citex}
  {\@citea\NAT@hyper@{%
     \NAT@nmfmt{\NAT@nm}%
     \hyper@natlinkbreak{\NAT@aysep\NAT@spacechar}{\@citeb\@extra@b@citeb}%
     \NAT@date}}
  {\@citea\NAT@nmfmt{\NAT@nm}%
   \NAT@aysep\NAT@spacechar\NAT@hyper@{\NAT@date}}{}{}

\patchcmd{\NAT@citex}
  {\@citea\NAT@hyper@{%
     \NAT@nmfmt{\NAT@nm}%
     \hyper@natlinkbreak{\NAT@spacechar\NAT@@open\if*#1*\else#1\NAT@spacechar\fi}%
       {\@citeb\@extra@b@citeb}%
     \NAT@date}}
  {\@citea\NAT@nmfmt{\NAT@nm}%
   \NAT@spacechar\NAT@@open\if*#1*\else#1\NAT@spacechar\fi\NAT@hyper@{\NAT@date}}
  {}{}
\makeatother

\makeatletter
\DeclareRobustCommand{\lowcase}[1]{\@lowcase#1\@nil}
\def\@lowcase#1\@nil{\if\relax#1\relax\else\MakeLowercase{#1}\fi}
\makeatother

\DeclareSymbolFont{UPM}{U}{eur}{m}{n}
\DeclareMathSymbol{\umu}{0}{UPM}{"16}
\let\oldumu=\umu
\renewcommand\umu{\ifmmode\oldumu\else\math{\oldumu}\fi}

\if\papercls \othercls

\else

\fi

\let\oldsim=\sim
\renewcommand\sim{\ifmmode\oldsim\else\math{\oldsim}\fi}
\let\oldpm=\pm
\renewcommand\pm{\ifmmode\oldpm\else\math{\oldpm}\fi}
\newcommand\by{\ifmmode\times\else\math{\times}\fi}
\newbox{\wdbox}
\renewcommand\c{\setbox\wdbox=\hbox{,}\hspace{\wd\wdbox}}
\renewcommand\i{\setbox\wdbox=\hbox{i}\hspace{\wd\wdbox}}

\newcount\timect
\newcount\hourct
\newcount\minct
\newcommand\now{\timect=\time \divide\timect by 60
         \hourct=\timect \multiply\hourct by 60
         \minct=\time \advance\minct by -\hourct
         \number\timect:\ifnum \minct < 10 0\fi\number\minct}

\catcode`@=11

\newcommand\comment[1]{}

\newcommand\commenton{\catcode`\%=14}

\renewcommand\math[1]{$#1$}
\newcommand\mathshifton{\catcode`\$=3}

\let\atab=&
\newcommand\atabon{\catcode`\&=4}

\let\oldmsp=\sp
\let\oldmsb=\sb
\def\sp#1{\ifmmode
           \oldmsp{#1}%
         \else\strut\raise.85ex\hbox{\scriptsize #1}\fi}
\def\sb#1{\ifmmode
           \oldmsb{#1}%
         \else\strut\raise-.54ex\hbox{\scriptsize #1}\fi}
\newbox\@sp
\newbox\@sb
\def\sbp#1#2{\ifmmode%
           \oldmsb{#1}\oldmsp{#2}%
         \else
           \setbox\@sb=\hbox{\sb{#1}}%
           \setbox\@sp=\hbox{\sp{#2}}%
           \rlap{\copy\@sb}\copy\@sp
           \ifdim \wd\@sb >\wd\@sp
             \hskip -\wd\@sp \hskip \wd\@sb
           \fi
        \fi}
\def\msp#1{\ifmmode
           \oldmsp{#1}
         \else \math{\oldmsp{#1}}\fi}
\def\msb#1{\ifmmode
           \oldmsb{#1}
         \else \math{\oldmsb{#1}}\fi}

\def\supon{\catcode`\^=7}

\def\subon{\catcode`\_=8}

\def\supsubon{\supon \subon}

\newcommand\actcharon{\catcode`\~=13}

\newcommand\paramon{\catcode`\#=6}

\comment{And now to turn us totally on and off...}

\newcommand\reservedcharson{ \commenton  \mathshifton  \atabon  \supsubon 
                             \actcharon  \paramon}

\catcode`@=12
\reservedcharson

\if\papercls \apjcls

\else

\fi

\newcommand\chisq{\ifmmode{\chi\sp{2}}\else\math{\chi\sp{2}}\fi}
\newcommand\redchisq{\ifmmode{ \chi\sp{2}\sb{\rm red}}
                    \else\math{\chi\sp{2}\sb{\rm red}}\fi}
\newcommand\Teq{\ifmmode{T\sb{\rm eq}}\else$T$\sb{eq}\fi}
\newcommand\mjup{\ifmmode{M\sb{\rm Jup}}\else$M$\sb{Jup}\fi}
\newcommand\rjup{\ifmmode{R\sb{\rm Jup}}\else$R$\sb{Jup}\fi}
\newcommand\msun{\ifmmode{M\sb{\odot}}\else$M\sb{\odot}$\fi}
\newcommand\rsun{\ifmmode{R\sb{\odot}}\else$R\sb{\odot}$\fi}
\newcommand\mearth{\ifmmode{M\sb{\oplus}}\else$M\sb{\oplus}$\fi}
\newcommand\rearth{\ifmmode{R\sb{\oplus}}\else$R\sb{\oplus}$\fi}
\usepackage{newtxtext,newtxmath} % to use same fonts for math

\newcommand{\sigsfr}{$\Sigma_{\mathrm{SFR}}$}
\newcommand{\relssfr}{$\log \mathrm{(sSFR/sSFR_{MS})}$}

\shorttitle{SFR Surface Density}
\shortauthors{Salim et al.}

\begin{document}

\title{\sigsfr--$M_*$ Diagram: A Valuable Galaxy Evolution Diagnostic to
  Complement (s)SFR--$M_*$ Diagrams}

%% AUTHOR/INSTITUTIONS FOR AASTEX6.1:
\author{Samir Salim} 
\affil{Department of Astronomy, Indiana University, Bloomington, IN 47405, USA}

\author{Sandro Tacchella}
\affil{Kavli Institute for Cosmology, University of Cambridge, Madingley
Road, Cambridge, CB3 0HA, UK}
\affil{Cavendish Laboratory, University of Cambridge, 19 JJ Thomson
Avenue, Cambridge, CB3 0HE, UK}

\author{Chandler Osborne} 
\affil{Department of Astronomy, Indiana University, Bloomington, IN 47405, USA}

\author{S.\ M.\ Faber} 
\affil{Department of Astronomy and Astrophysics, University of California, Santa Cruz, CA, 95064, USA}

\author{Janice C.\ Lee}
\affil{Space Telescope Science Institute, Baltimore, MD, 21218, USA}

\author{Sara L.\ Ellison} 
\affil{Department of Physics \& Astronomy, University of Victoria, Victoria, British Columbia, V8P 1A1, Canada}

%% AUTHOR/INSTITUTIONS FOR EMULATE APJ:
% \author{Patricio~E.~Cubillos\altaffilmark{1,2},
% Joseph~Harrington\altaffilmark{1},
% and
% Third~Author\altaffilmark{1}
% }
% \affil{\sp{1} Planetary Sciences Group, Department of
%               Physics, University of Central Florida, Orlando, FL 32816-2385\\
%        \sp{2} Space Research Institute, Austrian Academy of Sciences,
%               Schmiedlstrasse 6, A-8042, Graz, Austria}

\email{salims@indiana.edu}

% %% Extra info for aastex:
% \received{Yesterday}
% \revised{Today}
% \accepted{Tonight}
% \published{Tomorrow}
% \submitjournal{AASJournal}

% BRR

\begin{abstract}
  The specific star formation rate (sSFR) is commonly used to describe
  the level of galaxy star formation (SF) and to select quenched
  galaxies. However, being a relative measure of the young-to-old
  population, an ambiguity in its interpretation may arise because a
  small sSFR can be either because of a substantial previous mass
  build up, or because SF is low. We show, using large samples
  spanning $0<z<2$, that the normalization of SFR by the physical
  extent over which SF is taking place (i.e., SFR surface density,
  $\Sigma_{\mathrm{SFR}}$) overcomes this ambiguity. \sigsfr\ has a
  strong physical basis, being tied to the molecular gas density and
  the effectiveness of stellar feedback, so we propose \sigsfr--$M_*$
  as an important galaxy evolution diagram to complement (s)SFR--$M_*$
  diagrams. Using the \sigsfr--$M_*$ diagram we confirm the
  \citet{schiminovich07} result that the level of SF along the main
  sequence today is only weakly mass dependent---high-mass galaxies,
  despite their redder colors, are as active as blue, low-mass
  ones. At higher redshift, the slope of the ``\sigsfr main sequence''
  steepens, signaling the epoch of bulge build-up in massive
  galaxies. We also find that $\Sigma_{\mathrm{SFR}}$ based on the
  optical isophotal radius more cleanly selects both the starbursting
  and the spheroid-dominated (early-type) galaxies than sSFR. One
  implication of our analysis is that the assessment of the inside-out
  vs.\ outside-in quenching scenarios should consider both sSFR and
  $\Sigma_{\mathrm{SFR}}$ radial profiles, because ample SF may be
  present in bulges with low sSFR (red color).
\end{abstract}


\section{Introduction}\label{sec:intro}

The rate at which a galaxy is producing stars
\citep{schmidt59,salpeter59} is one of its most fundamental physical
properties. Integrated over cosmic time (and summing across the
progenitors in a merging tree), the star formation rate (SFR) gives
the total stellar mass of a galaxy, which is proportional to the
current stellar mass ($M_*$). The ability to constrain a galaxy's change
in SFR over time---its star formation (SF) history---lies at the heart
of the study of galaxy evolution.

The physical characterization of galaxies in terms of SFR and $M_*$ on
a massive statistical scale became possible with the advent of large
galaxy spectroscopic surveys, such as SDSS \citep{strauss02} and GAMA
\citep{driver11} at low redshift, and DEEP2 \citep{newman13}
and zCOSMOS \citep{lilly07} at higher redshifts, and has benefited from
the development of modern stellar population synthesis models (e.g.,
\citealt{bc03}) and new analysis techniques, including Bayesian
parameter estimation (e.g., \citealt{k03a,brinchmann04,s05}).

Construction of the diagrams that involve SFR and $M_*$ of a large
number of galaxies led to new insights and new conceptual frameworks,
such as the recognition of a relatively narrow range of SFR at a given
stellar mass---the star-forming main sequence---and its turn-off at
higher masses \citep{brinchmann04,bauer05,noeske07,s07,elbaz07}. For
galaxies on the main sequence, SFR to first order increases simply
because of the scale of a system, so it is useful to normalize it, the
usual choice being the stellar mass, yielding the specific SFR (sSFR,
\citealt{bothun82,tully82}). The SFR--$M_*$ diagrams, and the related
sSFR--$M_*$ diagrams \citep{guzman97,perez-gonzalez03}, have emerged
as a sort of equivalents of Herzsprung-Russel (HR) diagrams for
galaxies.

The specific SFR is commonly used as an indicator of the current level
of SF. High levels suggest a starburst---a galaxy experiencing an
increase in SFR compared to its baseline value, whereas low values, or
in some cases the upper limits, indicate a galaxy in which the SF has
been quenched. Various galaxy colors are also sometimes used as
indicators of SF activity. The specific SFR, like the color,
essentially represents the contrast between young and old stars, so it
combines the current SF state with its past history. Much effort in
recent years, both observational (e.g.,
\citealt{fang13,yesuf14,bluck14,woo15,pacifici16,barro17,martin17,rowlands18,belli19,moutard20,carnall20,yesuf22,tacchella22})
and theoretical (e.g.,
\citealt{ciambur13,sparre15,dubois16,feldmann16,tacchella16,weinberger17,weinberger18,behroozi19,dave19,donnari19,walters22})
has been invested to understand both the general SF history of a
galaxy while it is actively forming stars and the processes that lead
to its quenching. It would be beneficial for such efforts to use a
wide array of measures that illuminate relevant processes. The goal of
the present study is to point out the conceptual difference between
the two roles that sSFR is used for (current SF activity, including
the quenching, vs.\ the past SF history) and propose a way to separate
them.

Before proceeding, it is worthwhile to discuss the definition of
the verb `quenching' and adjective `quenched', since we will often
refer to them. As pointed out by \citet{belfiore18}, \citet{donnari19} and
others, there is no agreed upon definition of the verb
`quenching'. Definition of quenching is relatively unambiguous for
individual SF histories, especially the idealized ones. For them, the
quenching represents a downward change in SFR with respect to some
gradual overall trend. For example, \citet{martin07}, one of the first
studies on quenching, defines it as an exponential decline with a
variety of quenching rates ($e$-folding times 0.5 to 20 Gyr) following
a {\it constant} SFR. Given the difficulties in constraining the SF
histories of individual galaxies, and the fact that their forms in
reality contain complex details \citep{pandya17}, a definition that is
based on the properties of an ensemble of galaxies would be more
useful. Thus, the definition that \citet{belfiore18} and many other studies
use, and the one we will for most part use here, is that 'quenching'
refers to a {\it process that moves the galaxies below the main
  sequence}. For a galaxy of a certain mass, being below the main
sequence implies a different SF history than of other galaxies of the
same mass---one that currently has a lower SFR.

Using an adjective `quenched' to define a galaxy with {\it no} SF, can be
considered an {\it absolute} definition. Note that one can
alternatively consider a galaxy as quenched when its SFR no longer
contributes to the build-up of a galaxy on some relevant timescales,
such as the Hubble time at the epoch of observation. Such {\it
  relative} definition is closely tied to the sSFR
\citep{tacchella18}, since the inverse of sSFR represents the time
needed to double the stellar mass (modulo gas recycling). A galaxy
with log sSFR $=-10.1$ today will take a Hubble time to double its
mass, so one can think of it as having finished most of its assembly
and thus in a certain way quenched. If one uses this relative definiton of
being quenched, then sSFR is, by definition, the only way to identify
quenched galaxies. In this paper, we will assume the absolute definition
where quenched means no detectable SF. We acknowledge that depending
on the science question the relative definition may be more relevant.

`Quenching' implies that the final outcome of this process is a galaxy
wherein the SF is no longer present, i.e., it is quenched. However,
the galaxy below the main sequence with measurable SF is not yet
quenched, so it is useful to distinguish between a quenching and a
quenched galaxy. This distinction is embodied in the green valley
galaxies, which are often considered to be quenching, but are in any
case not yet quenched \citep{martin07,s14}. Therefore, we will
distinguish, when relevant, between galaxies undergoing quenching and
the ones fully quenched.

Returning to sSFR for the ensembles of galaxies, its ambiguous
interpretation can be illustrated using the sSFR--$M_*$ diagram of
present-day galaxies. Along the main sequence, the average sSFR
declines by $\sim$1 dex over the range of 3 dex in stellar mass. This
drop is not usually interpreted to mean that the massive galaxies are
collectively being quenched (or more strongly quenched) than the
low-mass ones. Rather, this tilt of the sSFR main sequence is equivalent to the
well-known fact that dwarf galaxies and late-type spirals are bluer in
optical color than early-type spiral galaxies such as M31
\citep[e.g.,][]{prugniel98}. Both the sSFR trend and the
(dust-corrected) color trend are the manifestations of the fact that
the SF histories of low and high-mass galaxies on the main sequence
differ, in the sense that the more massive galaxies have formed a
larger fraction of stellar mass earlier (galaxy ``downsizing'', 
\citealt{cowie96}), which is equivalent to saying that the massive
star-forming galaxies have older mean population ages.

The ideas regarding the mass-dependent SF histories of star-forming
galaxies date back to many decades ago
\citep{epstein64,tinsley80,sandage86}, with the subsequent empirical
confirmation that the colors or the sSFRs of star-forming galaxies are
mass-dependent being provided by
\citet{gavazzi96,gavazzi96b,k03b,brinchmann04}, and further confirmed
in simulations (e.g., \citealt{sparre15}). The confounding aspect of
sSFR for an ensemble of galaxies arises from the fact that in addition
to sSFR systematically decreasing across the main sequence even in the
absence of any process that quenches the SF, it obviously also
decreases if a galaxy quenches.

In this work, we wish to address this conceptual ambiguity of sSFR by
considering a measure that more closely represents the {\it current SF
  level} of a galaxy. We explore SFR normalized by the surface over
which SF takes place, i.e., the integrated SFR surface density
(\sigsfr), as such a quantity. The rationale for using the SFR surface
density can be illustrated by the following example. Consider two
galaxies having the same size (area over which SF takes place) and the
same SFR (or the number of HII regions). Intuitively we feel that we
should regard the two galaxies as having the same level of SF
activity, and the \sigsfr\ being the same for both galaxies supports
that notion. On the other hand, if one of the two galaxies has a more
massive disk or a more prominent bulge it would have a lower sSFR than
the other galaxy, despite their current level of SF being the same.

A popular alternative to sSFR in the context of SF level is the {\it
  relative} SFR, i.e., the logarithmic offset in SFR from the main
sequence \citep{schiminovich07,wuyts11,elbaz11}. (By definition, the
relative SFR and relative sSFR are identical.) The relative SFR has
been designated variously as \relssfr, $\Delta \log$ (s)SFR, or log
$\Delta_{\mathrm{SFR}}$. As a relative measure, its character is
different from that of either sSFR or \sigsfr. Its importance lies in
the fact that it is tied to the definition of quenching, as discussed
above, and its resulting practical use, e.g., for identifying
starbursts or quiescent galaxies, so we will analyze it alongside sSFR
and \sigsfr.

Although the SFR surface density is a familiar measure, it is
primarily used in the context of its relation with gas surface
density, i.e., the Kennicutt-Schmidt relation, either in integrated
(e.g., \citealt{kennicutt89}) or resolved sense (e.g.,
\citealt{kennicutt07}). Less often, \sigsfr\ is discussed as an
indicator of the level of SF activity---but rarely so in the context
of large samples of galaxies, notable exceptions being
\citet{schiminovich07} and \citet{wuyts11}. In any case, the
integrated \sigsfr\ is not as widespread a measure as sSFR and
warrants closer investigation. Thus, the aim of this paper is to
provide a comparative assessment of \sigsfr\ and sSFR, as well as the
SFR relative to the main sequence, and discuss the implication of
using \sigsfr, not just for integrated, but also for resolved galaxy
studies.

The paper is organized as follows. In Section \ref{sec:data} we
present our samples spanning three redshift ranges and associated
galaxy physical parameters obtained from the spectral energy
distribution (SED) fitting. In particular, in Section
\ref{ssec:sigsfr} we discuss the definition of \sigsfr\ and
differences arising from using different types of galaxy radii. In
Section \ref{sec:results} we present the comparative analysis of
sSFR--$M_*$ and \sigsfr---$M_*$ diagrams for different types of
galaxies at different redshifts. Section \ref{sec:disc} explores some
additional implications of the results and places them in the context
of previous work. We summarize our main findings in Section
\ref{sec:concl}.

%%%
\section{Sample and Data} \label{sec:data}
%%%

\subsection{Low Redshift Sample} \label{ssec:lowz}

Our principal sample comes from the GALEX-SDSS-WISE Legacy Catalog of
galaxy physical properties (GSWLC, \citealt{s16}). The catalog
contains galaxies corresponding to the SDSS main galaxy spectroscopic
survey ($0.01<z<0.30$ and $r<18$, \citealt{strauss02}) as long as they
were covered by GALEX \citep{martin05} UV surveys. GSWLC contains
three catalogs, according to the depth of GALEX imaging. In this work
we use the second release of the medium-deep catalog (GSWLC-M2,
\citealt{s18}), which offers a good combination of the UV depth
required to measure SFRs below the main sequence (transitional, green
valley galaxies) and a large sky coverage ($\sim$1/2 of SDSS area).

GSWLC-M2 contains 361 k galaxies at $0.01<z<0.30$, but we restrict our
analysis to 175 k galaxies at $z<0.1$, in order to ensure greater
reliability of the morphological data and higher UV detection
rates. The exclusion of type 1 AGNs and galaxies with poor SED fits
removes additional 2 k objects. GSWLC contains galaxies in the GALEX
footprint regardless of whether they were detected in the UV. Leaving
only the galaxies with the detections in either the far-UV or the
near-UV band (86 k) ensures more robust SFRs for the transitional
galaxies. A requirement that the galaxy has a measured isophotal size
and structural parameters (the data sources of which are based on SDSS
DR7, rather than DR10 used for GSWLC) results in the final
low-redshift sample of 77 k galaxies.

From this sample, for some analyses we focus on a subset of 29 k
main-sequence galaxies, which we define as the emission-line galaxies
that fall in the star-forming portion of the BPT diagram based on the
\citet{agostino21} modification of the \citet{k03c} demarcation
line. We require SNR$>2$ on the BPT emission lines, except SNR$>10$ on
H$\beta$.

We additionally match our final sample to the \citet{nair10} catalog of
visually determined Hubble types, which results in a smaller sample of
4 k galaxies (2 k main-sequence ones), and to the \citet{darg10} catalog
of merger pairs, the latter yielding 639 matches (240 main-sequence
ones). For merger pairs, we match our sample to each galaxy of the
pair. The match is typically the primary galaxy of the pair
($\sim$85\% of cases), because only about 1/4 of the secondaries have
SDSS spectra.

\subsection{High Redshift Sample}

Our high-redshift sample is taken from the CANDELS survey
\citep{grogin11,koekemoer11} of the GOODS-S field. We consider two
high-redshift windows: at $z \sim 1$ ($0.8 < z < 1.4$) and at
$z \sim 2$ ($1.5 < z < 2.5$). Redshifts are taken from \citet{guo13}
and are either photometric, or spectroscopic, when the latter are
available (see \citealt{santini15} for details). The sample is limited
to galaxies that appear in the \citet{vanderwel12} catalog of sizes
and structural parameters, and have $H$-band magnitude $< 24.5$. The
magnitude cut is beneficial to ensure the reliability of the
S\'{e}rsic indices and effective radii measurements. Our final sample
size is 1,905 galaxies at $z \sim 1$ and 1,550 galaxies at $z \sim 2$,
with 27\% and 15\% of redshifts being spectroscopic, respectively.

\subsection{Data}

GSWLC-2 provides stellar masses and dust-corrected SFRs derived from
the energy-balance SED fitting based on the photometry from GALEX,
SDSS and WISE. The models were generated and the fitting was
performed using CIGALE \citep{boquien19}. Models feature
two-component SF histories, as described in \citet{s16}, and flexible
dust attenuation curves described in \citet{s18}.

Stellar masses and dust-corrected SFRs for the high-redshift sample
are taken from \citet{osborne20}, and were derived using consistent SED
modeling as for the low redshift, with some adjustments needed to
account for the difference in look-back times. The principal
difference is that high-redshift SED fitting did not include dust
emission constraints, because IR observations are not available for
most of the sample.

All stellar masses and SFRs are based on \citet{bc03} stellar
population synthesis models and a Chabrier initial mass function
\citep{chabrier03}, and are given in units of Solar mass and Solar
mass per year, respectively.

Isophotal sizes for the low-redshift sample are taken from the
official SDSS DR7, and are based on 25 mag arcsec$^{-2}$ AB
isophotes. We use $r$-band minor and major axes, but find that other
bands produce equally robust results. Isophote at 25 mag arcsec$^{-2}$
AB in $r$ is roughly 0.7 mag deeper than the 25 mag arcsec$^{-2}$ Vega
isophote in $B$. Subsequent SDSS data releases omitted isophotal
sizes, claiming they were unreliable. We have found no evidence of any
issues. We also use effective (half-light) radii and S\'{e}rsic
indices from the \citet{meert15} catalog of structural parameters,
derived using GALFIT routine and assuming a single S\'{e}rsic profile.

\begin{figure*}[tb]
\centering
\includegraphics[width=\linewidth, clip]{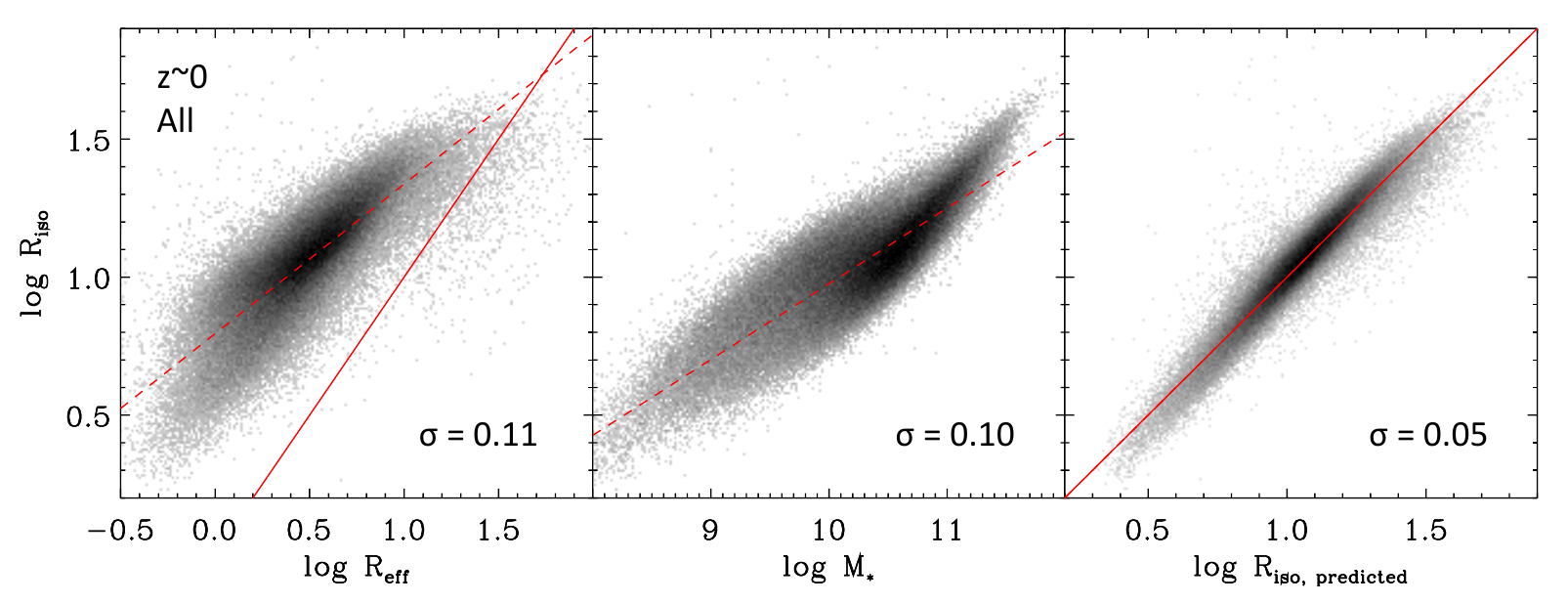}
\caption{Relationship between the isophotal radius (in kpc) and (a)
  the effective radius (also in kpc), (b) the stellar mass (in Solar
  masses), and (c) a combination of the isophotal size and stellar
  mass that minimizes the scatter (from left to right). Isophotal size
  is poorly correlated with either the effective size or the stellar
  mass, but is tightly correlated with the linear combination of the
  two (in logarithm), providing a way to precisely predict isophotal
  radius if it is not available (see Eq.\ \ref{eq:pred}). Standard
  deviations of the residuals around the best linear fit (dashed line)
  are given in the panels. Solid lines show 1:1 relation where
  applicable. All of the low-redshift samples in this and subsequent
  figures come from GALEX-SDSS-WISE Legacy Catalog (GSWLC-M2).}
\label{fig:riso_reff_mass_rpred}
\end{figure*}

Effective radii and S\'{e}rsic indices for the GOODS-S CANDELS
high-redshift sample are taken from \citet{vanderwel12}, and were also
derived using GALFIT. We use the effective radii and S\'{e}rsic
indices estimated from the $H$-band (F160W) imaging, which is roughly
comparable to to $r$ band at $z \sim 0$. We use the isophotal area
from the \citet{guo13} photometric catalog {\tt ISOAREAF}, also based
on F160W, to which we apply a redshift and a zero-point corrections
(Section \ref{ssec:pred}).

Visual morphological classification into Hubble types is taken from
the catalog of \citet{nair10}, based on SDSS images. The redshift
coverage of the catalog is the same as of our principal sample
($0.01<z<0.10$), but it contains far fewer objects (14 k) because it
targets brighter galaxies ($g<16$), and is based on DR4, which has
60\% of the area of DR7. We also utilize the sample of 3 k merger
pairs from \citep{darg10}, identified based on the Galaxy Zoo visual
classification of SDSS DR6 images (85\% of DR7 area). The parent
sample for this catalog has essentially the same redshift and
magnitude selections as our primary sample.

Emission-line data needed for the selection of low redshift main
sequence galaxies come from the MPA/JHU processing of SDSS DR7
spectra, which follows the methodology of \citet{tremonti04}. We 
verify some of our results using the MPA/JHU SFRs and stellar masses,
which were derived following the approach of \citet{brinchmann04}.

\subsection{SFR Surface Density (\sigsfr)}\label{ssec:sigsfr}

SFR surface density is defined as:

\begin{equation}
\Sigma_{\mathrm{SFR}} = \mathrm{SFR}/A
\end{equation}
where $A$ is the physical (not the projected) area of a galaxy. In the case of an effective radius, the area is taken to be:

\begin{equation}
A = 2 R_{{\mathrm{eff}, a}}^2\pi
\label{eq:area_eff}
\end{equation}
where $R_{{\mathrm {eff}, a}}$ corresponds to the semi-major axis ($a$) of the
ellipse containing $1/2$ of the galaxy light, and the factor of two is
meant to qualitatively take into account that the effective radius
includes half of the galaxy light. Note that many studies, including
\citet{meert15} that we use here, define
effective radius
as the geometric mean of the semi-major ($a$) and semi-minor ($b$)
axes, the so called circularized effective radius ($R_{\mathrm
  {eff}}$). In that case $ R_{\mathrm{eff}}^2\pi$ is not
the physical area but the projected area ($ab\pi$), and we can introduce the
projection factor ($f_{\mathrm{proj}}$) to go from the projected to
the physical area:

\begin{equation}
A = 2 f_{\mathrm{proj}}\ R_{\mathrm{eff}}^2\pi
\label{eq:area_eff}
\end{equation}
We determine $f_{\mathrm{proj}}$ empirically for the SDSS sample to be
1.26 (0.1 dex).

Similarly, for \sigsfr\ based on the isophotal ellipse we use:
\begin{equation}
A = f_{\mathrm{proj}}\,R_{\mathrm{iso}}^2\pi
\label{eq:area_iso}
\end{equation}
where the (circularized) isophotal radius ($R_{\mathrm{iso}}$) is the
geometric mean of the apparent isophotal major and minor axes (given
as {\tt isoA} and {\tt isoB} in SDSS DR7) and converted from angular
to physical size using scale $s(z)$:
\begin{equation}
R_{\mathrm{iso}}=0.5 s(z) \sqrt{{\tt isoA}\, {\tt isoB}}
\label{eq:riso}
\end{equation}
The rationale for using the circularized sizes is that the linear extent
($a$) is less robustly constrained than $\sqrt{ab}$.

For galaxies without well-defined disks, such as the irregular
galaxies and post-mergers, the isophotal ellipse may not be the most
accurate way to establish the area. In such cases one can obtain the
area as the sum of the area of the pixels that lie above some surface
brightness threshold. Photometry software SExtractor \citep{bertin96}
provides such measurement as {\tt ISOAREAF} and {\tt ISOAREA}. This
area should replace $R_{\mathrm{iso}}^2\pi$ in Equation \ref{eq:riso}.

\subsection{Obtaining isophotal size from effective size and stellar
  mass} \label{ssec:pred}

As we will show, \sigsfr is a more precise indicator of SF activity if
derived using the isophotal size (or area), rather than the effective
(half-light) size (Section \ref{ssec:reff}). However, the isophotal
size or area is often not reported in photometric catalogs. Or, if it
is, it may be based on different thresholds. To overcome these
practical issues, we devise a transformation from effective to
isophotal radius, which we calibrate using our low-redshift, SDSS
sample, and validate using CANDELS $0.8<z<2.5$ data.

The left panel of Figure \ref{fig:riso_reff_mass_rpred} shows that the
correlation between isophotal and effective physical radii is strongly
non-linear and has a
non-uniform scatter, suggesting that a direct conversion from
effective to isophotal radii would be neither accurate nor
precise. Fortunately, there is a way to improve the matters using the
stellar mass. The middle panel of Figure
\ref{fig:riso_reff_mass_rpred} shows that the isophotal radius is also
correlated with the stellar mass, though not much better than with
respect to the effective radius, with the formal scatter around the best fit being
similar---0.10 vs.\ 0.11 dex. However, it turns out that the
$R_{\mathrm{iso}}$ is very well correlated with the combination of the
effective radius and stellar mass, and with a scatter of just
0.05 dex:

\begin{equation}
\log R_{\mathrm{iso}} =0.188\log M_* + 0.333\log R_{\mathrm{eff}}-1.037,
\label{eq:pred}
\end{equation}
where sizes are in kpc and the stellar mass is in units of Solar
mass. Isophotal size corresponds to 25 mag AB in $r$. The coefficients
in Equation \ref{eq:pred} were determined from a linear regression. In
other words, stellar mass, isophotal radius and effective radius form
a relatively tight 3D plane that contains both early and late type
galaxies. The right panel of Figure \ref{fig:riso_reff_mass_rpred}
shows $R_{\mathrm{iso}}$ predicted from this relation to be reasonably
unbiased. We tested expanding the calibration to include additional
parameters (the S\'{e}rsic index or SFR), but further gains were very
small ($\lesssim$2\% reduction in scatter).

We expect the observed isophotal size to be affected by the
cosmological surface brightness dimming. Indeed, when applying Equation
\ref{eq:pred} to CANDELS, we find the difference between the predicted
and observed isophotal radii to be strongly redshift dependent. This
allows us to construct a relation to use to correct the observed isophotal
size of a galaxy at redshift $z$ to its value at $z \sim 0$:

\begin{equation}
\log R_{\mathrm{iso}} =\log R_{\mathrm{iso}}^{\mathrm{obs}} + 0.129\, z
\label{eq:zcorr}
\end{equation}
Note that the relation has not been tested at $z>2.5$.

Finally, we check the validity of our calibration for higher redshifts
by comparing the predictions from Equation \ref{eq:pred} to the
actual, redshift-corrected isophotal sizes in CANDELS
($0.8<z<2.5$). The scatter between real and predicted sizes is 0.09
dex, in contrast to 0.17 dex between the isophotal and effective
radii. Comparison reveals a small zero point offset (0.033 dex),
arising from different surface brightness thresholds used in SDSS and
CANDELS. 

One potential concern is that some of the reduction in scatter in the
\sigsfr\ main sequence when using isophotal sizes derived from
Equation \ref{eq:pred} compared to effective sizes could be the
product of the covariances introduced by the calibration itself. To
test this, we compare the width of the SDSS \sigsfr\ main sequence when
using the isophotal sizes derived from Equation \ref{eq:pred} and
using the real \sigsfr. The latter is smaller (0.31 dex vs.\ 0.28
dex), suggesting that covariances do not affect it.

\begin{figure*}[tb]
\centering
\includegraphics[width=\linewidth, clip]{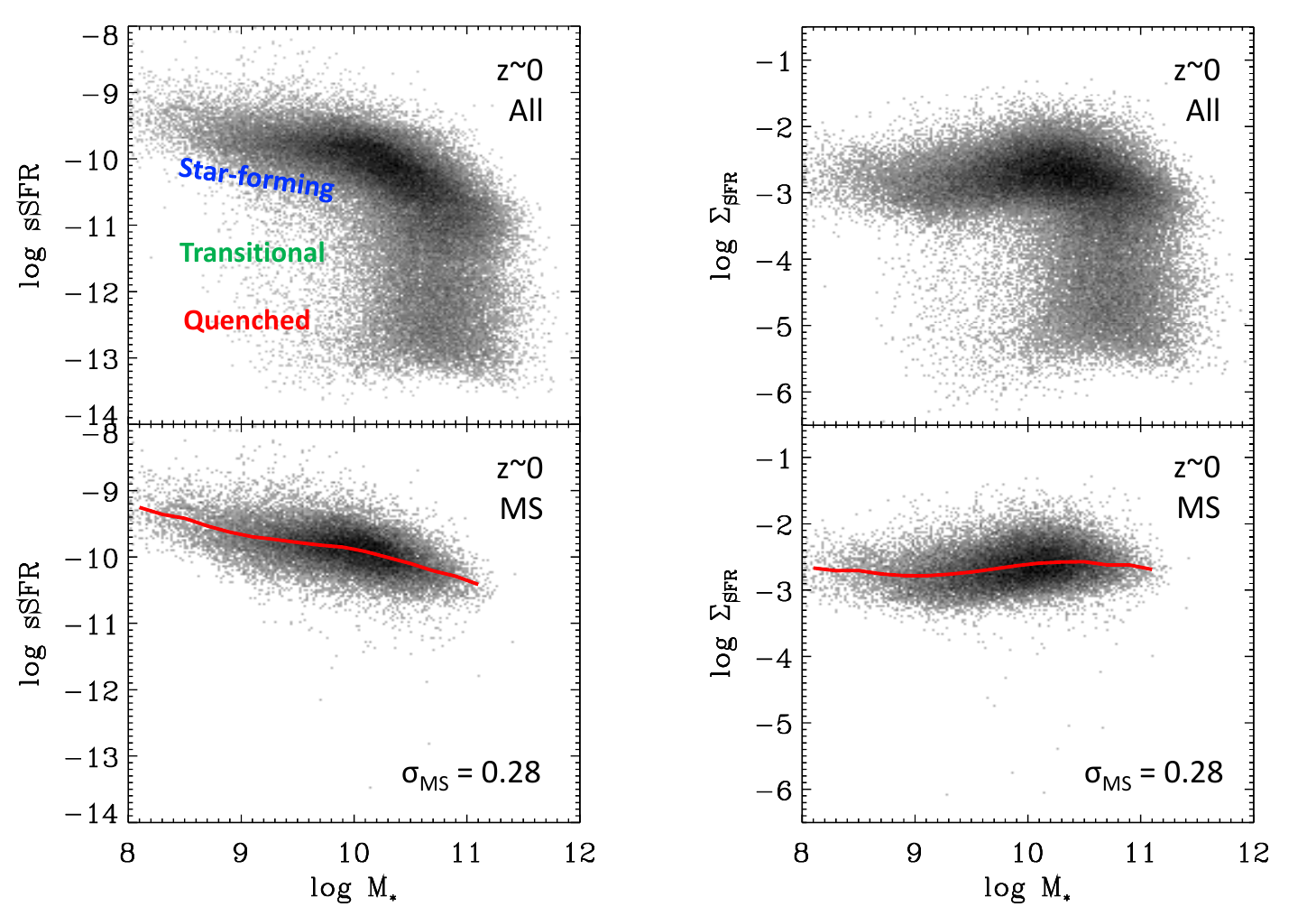}
\caption{Low-redshift galaxy samples shown on the sSFR--$M_*$ (left)
  and \sigsfr---$M_*$ diagrams (right). Lower panels in both figures
  show the subset of galaxies that form the main sequence. The
  character of the main sequence changes when going from sSFR to
  \sigsfr\ --- it becomes more flat, while remaining tight (width, in
  dex, given in panels). Once the measure of SF activity is decoupled
  from the past SF history by using \sigsfr, the SF level is not
  strongly mass dependent.}
\label{fig:ssig_sfr_mass}
\end{figure*}

\begin{figure*}[tb]
\centering
\includegraphics[width=\linewidth, clip]{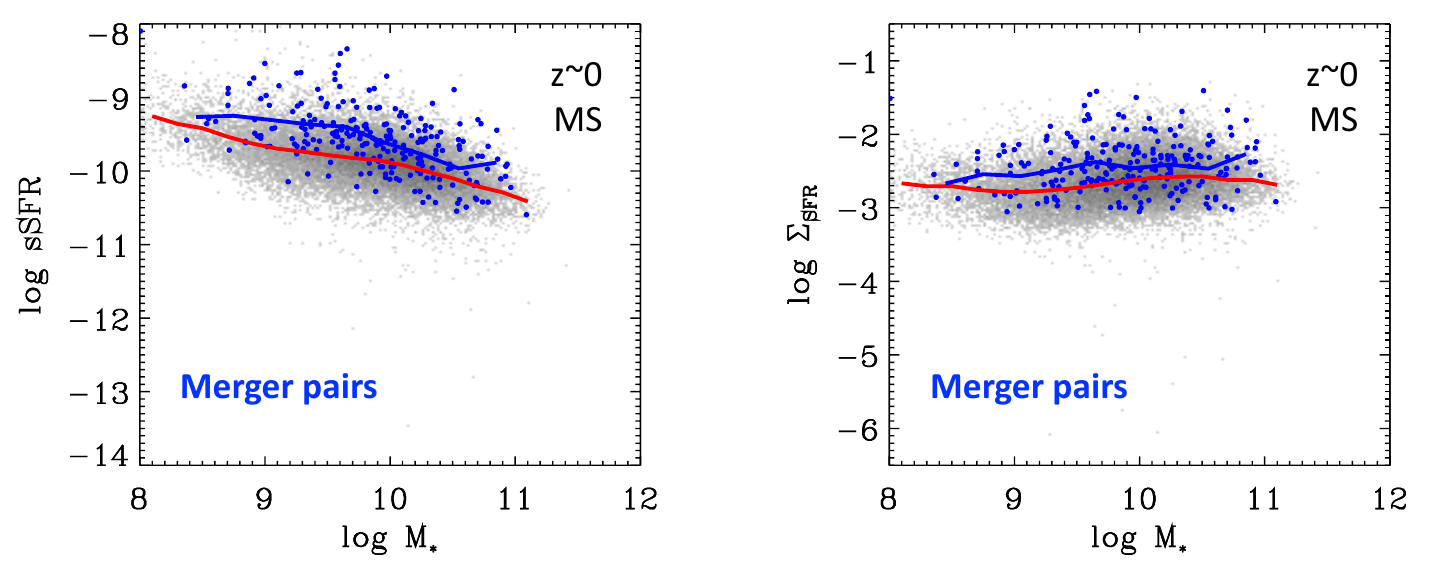}
\caption{Main-sequence merger pairs (interacting galaxies)
  over-plotted on the general low-redshift main sequence
  population. Same galaxies are shown on the sSFR--$M_*$ (left) and
  \sigsfr---$M_*$ diagrams (right). Merger pairs serve as proxies for
  starbursts and come from \citet{darg10}. Merging galaxies have a
  similar average offset (blue line) from the ridge of the overall
  main sequence (red line) in both cases; however, because there is no
  mass dependence, selecting starbursts by SFR surface density is
  obviously much cleaner than based on sSFR.}
\label{fig:ssig_sfr_mass_darg}
\end{figure*}

%%%
\section{Results} \label{sec:results}
%%%

In this section we explore, via sSFR--$M_*$ and \sigsfr--$M_*$
diagrams, the differences between sSFR and \sigsfr\ as indicators of
the degree of SF activity or quiescence. The analysis first focuses on
the low-redshift sample and its various subsamples before turning to
higher redshifts and the comparison to low redshift.

\subsection{Low-redshift main sequence}\label{ssec:lowzms}

The left panels of Figure \ref{fig:ssig_sfr_mass} show the sSFR--$M_*$
diagram for the low-redshift sample. The upper panel contains all the
galaxies in the sample, spanning 3.5 dex in stellar mass
($8<\log M_*<11.5$) and some 5 dex in sSFR
($-13.5<\log \mathrm{sSFR}<-8.5$). Higher-mass star-forming galaxies
tend not to reach sSFR values that are as high as sSFRs of low-mass
galaxies. This tilt of the main sequence can be seen more clearly in
the lower left panel, which shows only the galaxies selected as
star-forming in the BPT diagram. The tilt in the sSFR--$M_*$ main
sequence, i.e., that $\alpha<1$ in SFR $\propto M_*^{\alpha}$, is a
robust feature that exists irrespective of the choice of SFR indicator
(Appendix \ref{app:other_ms}).

As discussed in Section \ref{sec:intro}, the tilt of the main sequence
is an indication of mass-dependent SF histories and not of any change
that would suggest quenching (a downward departure from an overall
trend in SF history). However, an actual quenching would also lower
sSFR, introducing an ambiguity in interpretation of a sSFR. \sigsfr\
does not have this ambiguity and it should tell us about the current
level of SF activity. Therefore, we now take a look at the lower
panels of Figure \ref{fig:ssig_sfr_mass}, allowing us to contrast the
appearance of the main sequence in sSFR--$M_*$ and in \sigsfr---$M_*$
diagrams. The two panels span the same dynamic range (6 dex) in the
$y$ direction. The surface area used to normalize the SFR is based on
isophotal sizes. This choice will be discussed in Section
\ref{ssec:reff}. Most notably, we see that the main sequence no longer
has a downward tilt, and is quite flat (standard deviation of the
mass-binned averages of \sigsfr\ is 0.07 dex and the maximum amplitude
of binned averages is 0.21 dex). There are possible breaks at
$\log M_*=9.0$ and 10.4, which, by the way, are not at the same exact
masses as the breaks in the sSFR--$M_*$ main sequence ($\log M_*=9.1$
and 10.0). These subtle features aside, the first conclusion we draw in
this study is that the SF level of the present-day star-forming
galaxies is remarkably constant across the stellar mass.

The defining feature of the sSFR--$M_*$ (or, equivalently, SFR--$M_*$)
main sequence is that it is relatively tight. We find the width of
both the \sigsfr---$M_*$ and the sSFR--$M_*$ main sequence to be the
same (average of standard deviations in mass bins is 0.28 dex). This
is despite the fact that the measurement error of \sigsfr\ must be
greater than on sSFR because it includes the error on galaxy size.

\begin{figure*}[tb]
\centering
\includegraphics[width=\linewidth, clip]{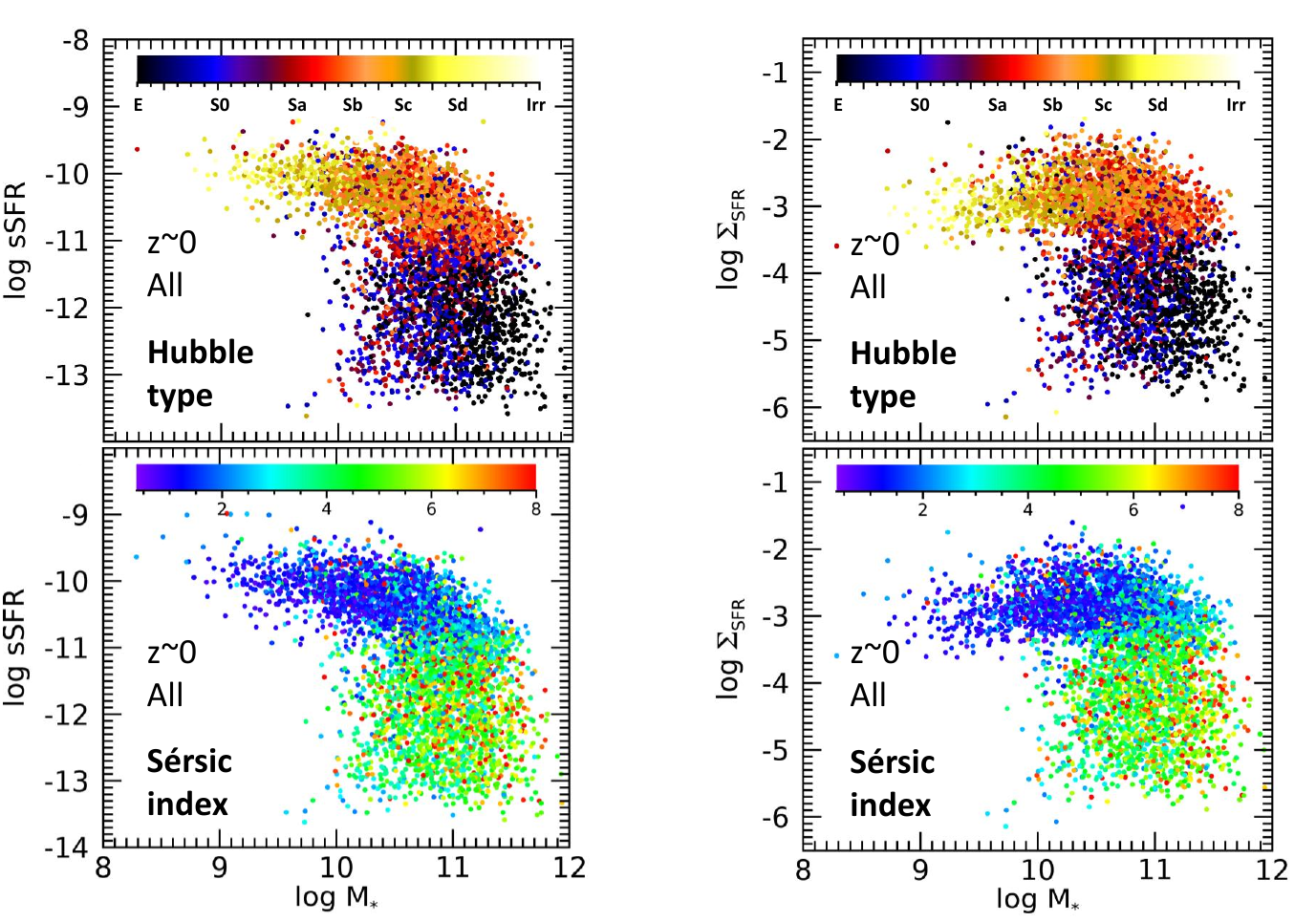}
\caption{Subsample of low-redshift galaxies with Hubble types from
  \citet{nair10}, shown on the  sSFR--$M_*$ (left) and \sigsfr---$M_*$
  diagrams (right). Upper panels are color-coded by the Hubble type,
  and lower panels by the S\'{e}rsic index (from \citealt{meert15}). Galaxies
  on the main sequence are mostly later types and have smaller S\'{e}rsic
  indices. SFR surface density provides a cleaner demarcation between
  early and late types than sSFR.}
\label{fig:ssig_sfr_mass_morph}
\end{figure*}

\subsection{Low-redshift starbursts} 

Next we explore how well does \sigsfr\ identify starburst galaxies
compared to sSFR. To answer this, we need a sample of starbursts that
is not identified using any SFR-related measure, because such
reasoning would obviously be circular. That includes selecting by the
the birth parameter $b=\mathrm{SFR}/$<SFR>, which is equivalent to
an sSFR selection for a fixed galaxy age. Instead, here we utilize the fact
that galaxy interactions can lead to an enhancement of SF and
therefore such galaxies are more likely to be starbursting (e.g.,
\citealt{barton00,patton13}).

In Figure \ref{fig:ssig_sfr_mass_darg} we overplot on the general
main-sequence sample the subsample of such galaxies identified
visually in \citet{darg10} as merging pairs. We include all pairs
regardless of the merging stage (separated, interacting, and
approaching post-merger). The great majority of this sample are
identified as being in the interacting stage. The left panel shows the
sSFR--$M_*$ diagram. We see that the merging galaxies are indeed
offset from the main sequence, with the mass-binned average difference
of 0.29 dex (a factor of 2.0 enhancement, the same as found in
\citealt{osborne20} for galaxies out to $z \sim 2$, and in
\citealt{patton13} for pair separations $<30$ kpc). The plot shows
that what makes a starburst is the relative enhancement in
SF---selecting starbursts on some fixed sSFR value is clearly not
justified. The offset between the interacting galaxy main sequence and
the overall main sequence is not strongly mass dependent. From this it
is immediately clear why the {\it relative} (s)SFR provides a much
more meaningful way to select starbursts than any sSFR cut. The width
of the interacting galaxy main sequence (0.36 dex) is not much larger
than of the general main sequence (0.28 dex), suggesting a relatively
uniform degree of SFR enhancement, having a standard deviation of 0.22
dex. Note that this standard deviation of SFR enhancement is likely
partially suppressed by the SFR averaging timescale that our measure
of SFR employs ($\sim$ 100 Myr) being longer than the starburst
timescale ($\sim$50 Myr, \citealt{wuyts09,tacchella20}).

% The st dev is partially suppressed by UV averaging timescale. In Ha
% the general MS is 0.38 dex wide comp w/ 0.51 dex for interacting. So
% stdev of enhancement is 0.33 dex which agrees perfectly with a random
% sampling over 750 Myr of a Gaussian burst with an amplitude of 1.2 dex
% and 1 sigma width of 50 Myr, as in Wuyts09 Fig 1.

Turning the attention to the right panel of Figure
\ref{fig:ssig_sfr_mass_darg}, we see that the \sigsfr\ main sequence
of interacting galaxies is flat, and that it shows a similar offset
with respect to the general main sequence, with the mass-binned
average difference of 0.21 dex (a 60\% enhancement). The width of the
interacting galaxy \sigsfr main sequence is even more similar to the
overall \sigsfr\ main sequence (0.32 dex vs.\ 0.28 dex), implying a
standard deviation of the enhancement in \sigsfr\ of just 0.13
dex. From the flatness of interacting galaxy \sigsfr\ main sequence we
conclude that a selectivity of \sigsfr\ to starbursts must be at least
as good as using the relative SFR.  We confirm this more directly by
comparing \relssfr\ to log \sigsfr, and finding that \citet{darg10}
star-forming mergers form an even sharper lower boundary in SFR
surface density (at log \sigsfr\ $=-3.0$) than in the relative
SFR. Selection of starbursts using a threshold in \sigsfr\ is not a
new thing \citep{kennicutt12}, but its preference to other methods has
not been universally recognized.

\begin{figure*}[tb]
\centering
\includegraphics[width=\linewidth, clip]{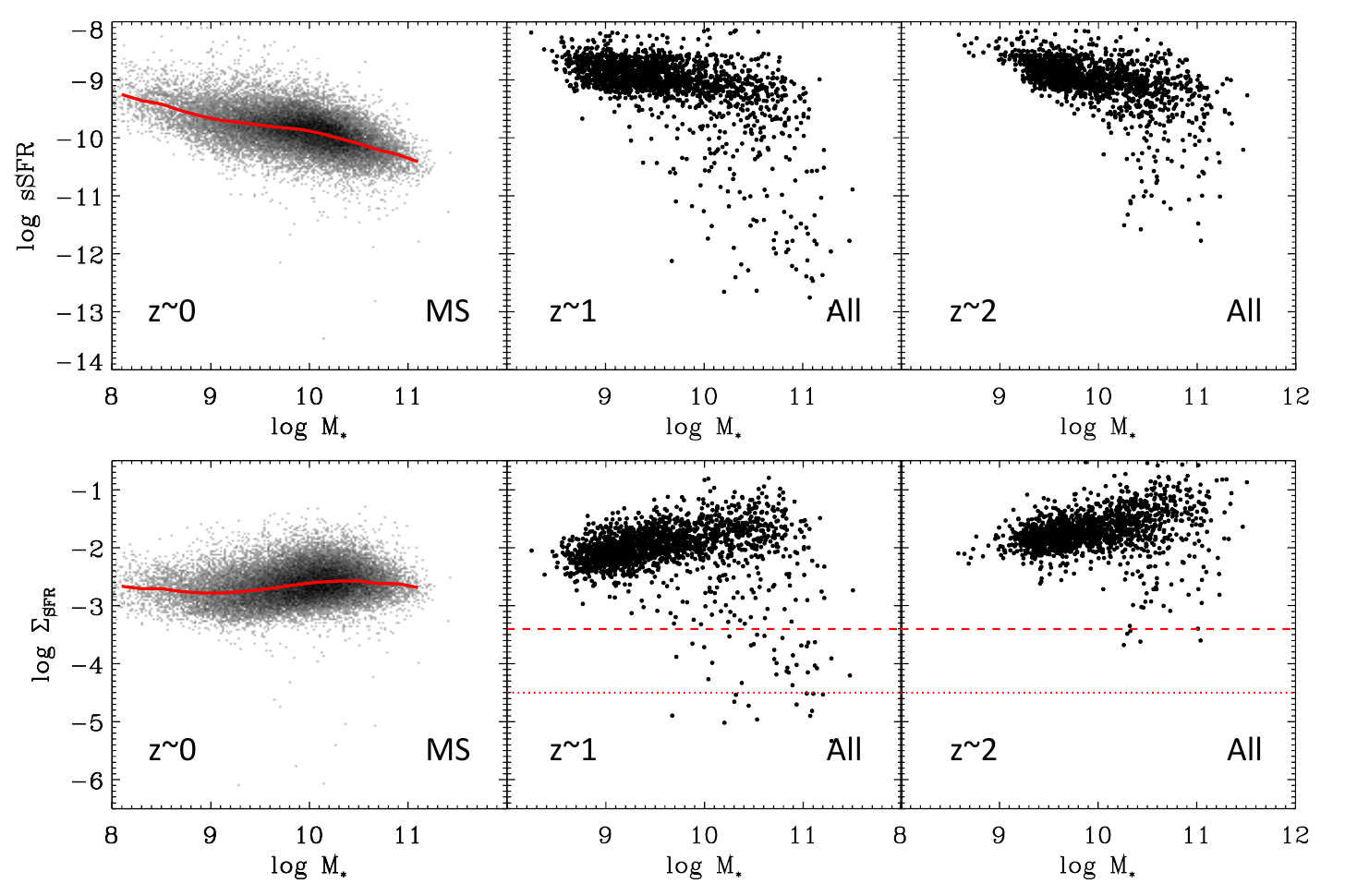}
\caption{Evolution of galaxies in sSFR--$M_*$ (upper panels) and 
  \sigsfr---$M_*$ diagrams (lower panels). For low-redshift samples we
  show only the main-sequence galaxies. In sSFR--$M_*$ diagrams the
  slope of the main sequence stays more or less the same, whereas
  \sigsfr---$M_*$ diagrams reveal a change---there was a higher SF
  activity of massive galaxies compared to less massive ones in the
  past, consistent with a bulge build-up of those systems. Dashed line
  in lower panels signifies the upper boundary of the low-redshift
  transition region. At higher redshift this boundary is higher and
  would have encompassed today's main sequence galaxies. The dotted
  line represents the threshold for full quenching, which is
  independent of the redshift.}
\label{fig:ssig_sfr_mass_z0_z1_z2}
\end{figure*}

\subsection{Low-redshift transitional/quenched galaxies} 

Before becoming fully quenched, the galaxies must transition, so that
their sSFRs are significantly lower than on the main sequence, but are
not yet entirely devoid of SF. It should be noted that observationally
establishing a complete absence of SF is challenging. Although not
formally having a SFR of zero (which the SF history parameterization
used in the SED fitting does not even allow), the galaxies with log
sSFR $<-12$ typically show no evidence of star-forming regions in UV
images and can be considered as fully quenched for all practical
purposes.

We next ask how well does \sigsfr\ separate the actively star-forming
galaxies from the ones that have or are experiencing quenching, i.e.,
the transitional galaxies. As in the case of starbursts, we cannot
define the quenching galaxies using any of the measures that we wish
to evaluate (sSFR, relative SFR and \sigsfr). Instead, we will take
advantage of the fact that the transitional/quenching galaxies, unlike
the ones that are not quenched, all have prominent spheroidal
components.  Qualitatively, this means that we expect the transitional
galaxies to be dominated by early-type galaxies (ellipticals,
lenticulars and early-type spirals). Quantitatively, we expect the
profiles of the galaxies to be more highly concentrated, as reflected
in their S\'{e}rsic indices being higher.

In Figure \ref{fig:ssig_sfr_mass_morph} we again show sSFR--$M_*$
(left panels) and \sigsfr---$M_*$ diagrams (right panels) of galaxies
matched to \citet{nair10} catalog and color code them by the Hubble
types provided in that catalog (upper panels), and by the S\'{e}rsic
indices from \citet{meert15} (lower panels). A clear trend is present
on the sSFR--$M_*$ plots---the main sequence is dominated by late-type
spirals (Sd at $\log M_*<9.8$, Sc and Sb types above that mass),
whereas the region below the main sequence is dominated by ellipticals
and S0s, with the former being more dominant at higher
mass. Interestingly, the main-sequence spirals can be as massive as
any but the most massive of the early-type galaxies. A trend similar
to that with the Hubble type is seen with respect to the S\'{e}rsic
index, where we see a significant increase in a typical S\'{e}rsic
index starting just below the main sequence. Note, however, that at
any point in the sSFR--$M_*$ plane there exists a range of S\'{e}rsic
indices and Hubble types, i.e., the trends are there, but are not very
tight. This confirms the finding of \citet{wuyts11} that the galaxies
are not just a two-parameter (SFR, $M_*$) family.

We can see that the ``threshold'' for the morphological transition in
sSFR--$M_*$ is somewhat tilted, following the tilt of the main
sequence. This suggests that, as in the case of starbursts, it is the
relative SFR that provides a cleaner morphological distinction than
sSFR, as previously pointed out by \citet{wuyts11}. On the other hand,
in \sigsfr---$M_*$ (Figure \ref{fig:ssig_sfr_mass_morph}, upper right
panel) the transition in Hubble types is essentially flat, i.e., it
happens at a fixed \sigsfr. This transition in \sigsfr\ is as sharp as
it is for the relative SFR. We confirm this quantitatively by
determining the interval over which the fraction of early-type
galaxies ($T \leq 0$) increase from the level typical for the main
sequence (12\%) to the level typical among the quiescent galaxies
(93\%).  This interval is 1.0 dex for both log \sigsfr\ and \relssfr,
compared to 1.2 dex for log sSFR. We also fitted a logistic function
to the fraction of ETGs vs.\ the parameter, and the slope is steepest
with respect to \sigsfr. Similar results are found regarding
how sharply the fraction of galaxies with high S\'{e}rsic index
($n>2.5$, corresponding to classical bulges, \citealt{drory07}) rises
as a function of decreasing sSFR, relative SFR, or \sigsfr. To
conclude, \sigsfr\ produces a cleaner separation between late and
early-type galaxies than sSFR, and is as clean as the relative SFR.

% Furthermore, the fraction of $n>2.5$ galaxies plateaus for the low
% (relative) sSFR at 95\%, whereas it reaches 100\% for the lowest
% \sigsfr, further attesting to its power.

\begin{figure*}[tb]
\centering
\includegraphics[width=\linewidth, clip]{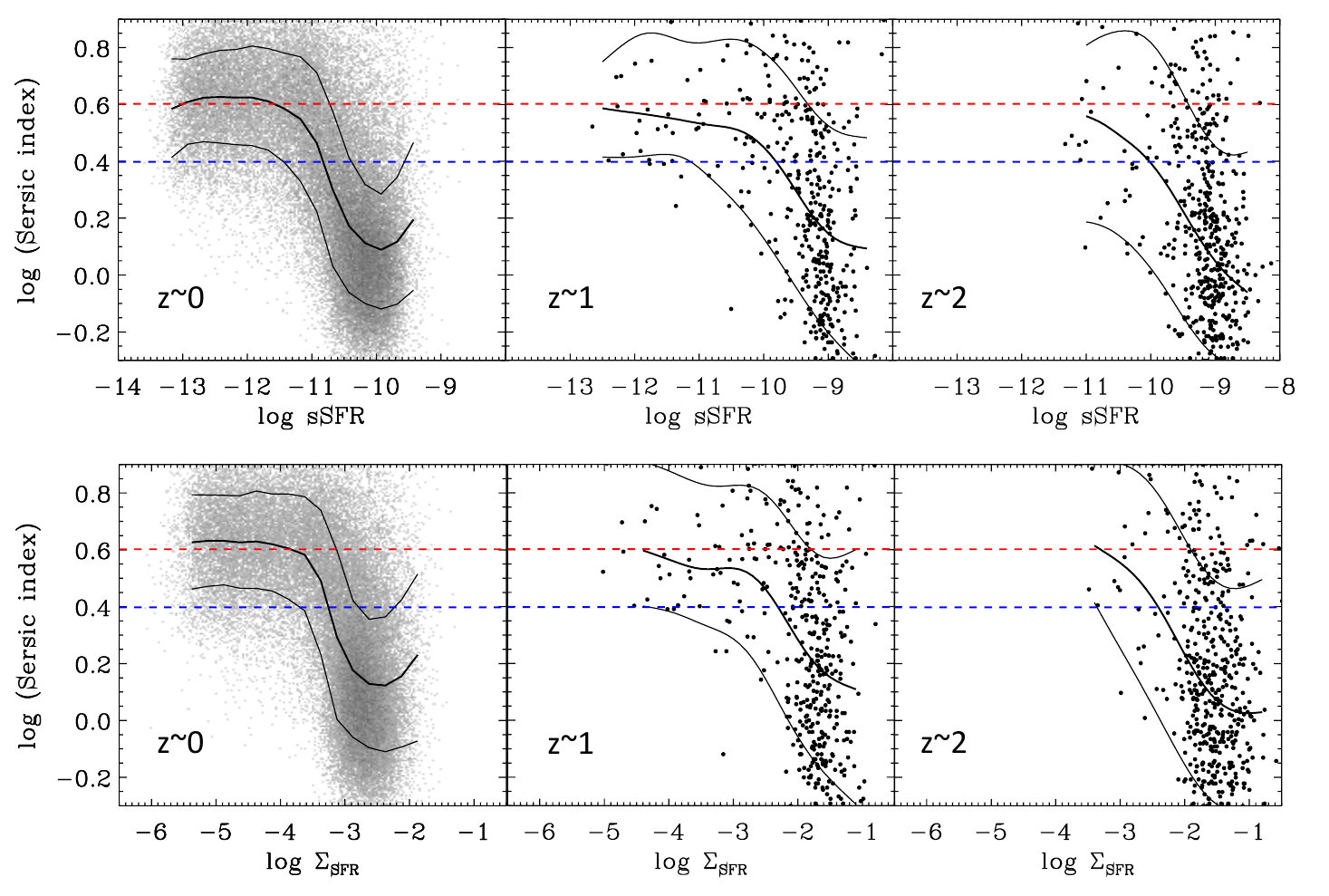}
\caption{Trends of the S\'{e}rsic index as a function of sSFR (upper
  panels) and \sigsfr\ (lower panels), at different redshifts. The
  galaxies with today's values of sSFR or \sigsfr\ characteristic of
  the main sequence would be found below the main sequence at high
  redshift, and like the transiting galaxies today, would have, on
  average, a high S\'{e}rsic index. This suggests that morphological
  transformation is associated with a relative suppression of sSFR or
  \sigsfr, but to values which would still be considered quite high for
  today's norms. The red dashed line corresponds to $n=4$ de
  Vaucouleurs' profile, and the red line corresponds to $n=2.5$, an
  approximate dividing line between the galaxies with and without a
  central spheroid. Shown are the galaxies with $10< \log M_* <
  11$. Curves represent the 16 percentile, the mean, and the 84
  percentile in each $x$-axis bin. Curves at $z \sim 1$ and $\sim 2$
  are smoothed to mitigate low numbers of objects.}
\label{fig:nsersic_ssig_sfr_z0_z1_z2}
\end{figure*}

\subsection{High-redshift main sequence}

%% Slope calc from fits to binned averages in ssig_sfr_mass_ms3.xlsx 

Our attention is now turned to higher redshift samples and the
evolution between $z \sim 2$ and $z \sim 0$. In the upper panels of
Figure \ref{fig:ssig_sfr_mass_z0_z1_z2} we are showing the sSFR-$M_*$
diagrams at $z\sim 0$, 1 and 2. We see that the fraction of
high-redshift ($z\sim 1$ and 2) galaxies below the main sequence is
much smaller than in SDSS, so for clearer comparison we only show the
main-sequence sample for low redshift. We first note that there is not
much evolution in sSFR-$M_*$ between $z\sim 2$ and $z\sim 1$---the
slope and the normalization of the main sequence are
similar. Actually, the slope of the main sequence is not very
different at low redshift either ($-0.34$ at $z\sim 0$ compared to
$-0.27$ at $z\sim 1$ over the same mass range, $9<\log M_*<11$), as
noted already in \citet{noeske07}. However, as expected, the
normalization today is significantly lower---0.9 dex with respect to
$z\sim 1$. So, based on the invariant slopes of the main sequence in
sSFR (or, alternatively, just SFR) one might conclude that SF is less
active at all masses by a similar degree.

The picture looks different with \sigsfr\ replacing sSFR (lower panels
of Figure \ref{fig:ssig_sfr_mass_z0_z1_z2}). Again, there seems to be
little evolution between $z\sim 2$ and $z\sim 1$ in terms of the slope
and the normalization of the \sigsfr\ main sequence. However, unlike
the case of sSFR-$M_*$, we see a change between low and high redshift,
not only in the normalization, but also in the slope of the main
sequence. The slope is slightly positive at $z\sim 0$ (slope 0.11),
whereas it is 0.25 at $z\sim 1$. So, based on the main sequence in
\sigsfr\ we conclude that the level of SF activity since the ``cosmic
noon'' has dropped more for massive galaxies ($9\times$ for
$10.6< \log M_*<11.0$) than for low-mass ones ($5\times$ for
$9.0< \log M_*<9.4$).

\subsection{High-redshift transitional/quenched galaxies}\label{ssec:highz}

Here we address a question: can the distinction
between a non-quenched and a transitional galaxy be based on a
redshift-invariant parameter? We know that sSFR does not provide this,
because the normalization of the main sequence changes, and the lower
panels of Figure \ref{fig:ssig_sfr_mass_z0_z1_z2} show that neither
does \sigsfr---the \sigsfr\ main sequence also changes in redshift,
as discussed in the preceding section. In other words, at a given
stellar mass a typical high-redshift galaxy will have both the higher
SFR and the higher SFR per surface area compared to a present-day
galaxy.

To further explore if it is justified to consider a galaxy with the
same $M_*$ and the same SFR (or \sigsfr) as quenching at one redshift
and not quenching at another, we again look at the structural
properties. Figure \ref{fig:nsersic_ssig_sfr_z0_z1_z2} presents, for
the same three redshift bins, the S\'{e}rsic index as a function of
sSFR (upper panels) and \sigsfr\ (lower panels). At $z\sim 0$, the
average S\'{e}rsic index for a galaxy at log sSFR $=-10.2$ (main
sequence at that redshift) is $n=1.3$---an exponential disk. On the
other hand, the average S\'{e}rsic index at that same sSFR at
$z\sim 1$ is $n=3.0$, typical of early-type galaxies today. A similar
result is obtained if considering \sigsfr. From this we confirm the
\cite{wuyts11} conclusion that the transitional/quenching status is
indicated by the SFR (or in our case also \sigsfr) relative to the
main sequence at that redshift, and not by any absolute threshold in
either sSFR or \sigsfr.

Whereas the criterion for the onset of quenching remains tied to the
position with respect to the main sequence for either the (s)SFR or
\sigsfr\ main sequence, the achievement of full quiescence is still
meaningful as a fixed, redshift-independent threshold in \sigsfr, but
not in sSFR or relative SFR. The reason for this is again that unlike
sSFR, \sigsfr\ is not the young-to-old population contrast, but an
absolute measure of the young population. Similar ambiguity regarding
the threshold for full quiescence would apply to a color, or H$\alpha$
equivalent width, as both are contrasting the young and old
population. Locally, the bottom boundary of the transitional region
appears to be around log \sigsfr\ $=-4.5$. Establishing the lowest
\sigsfr\ for transitional galaxies is difficult because of the
difficulties involved with measuring very low levels of SF, so we
consider this threshold as provisional. If we take this threshold to
be redshift independent, Figure \ref{fig:ssig_sfr_mass_z0_z1_z2} shows
that very few galaxies at $z \sim 1$ and no galaxy at $z \sim 2$ fall
below it. It should be noted however that constraining these low
\sigsfr\ values, especially at high redshift is challenging and
sensitive to the assumptions regarding the SF histories used in the
SED fitting, so it is difficult to know for sure if any of them fall
below the log \sigsfr\ $=-4.5$ threshold.

We should also point out that the \sigsfr\ radial profiles of galaxies
typically decline towards the outskirts (e.g., \citealt{gildepaz07}),
which means that any threshold that we want to attach to full
quiescence will depend on the size used for \sigsfr. If we were to use
the effective radii instead of isophotal, the proposed threshold for
quiescence would be $\sim$1 dex higher (log \sigsfr\
$=-3.5$). Similarly, the \sigsfr\ defining the lower envelope of the
main sequence (the threshold for the onset of transitional region)
will de different depending on the size of the aperture used to obtain
\sigsfr.

There are several additional things one can infer from Figure
\ref{fig:nsersic_ssig_sfr_z0_z1_z2}. 

\begin{enumerate}

\itemsep -0.5\parsep 

\item At all redshifts, galaxies with high S\'{e}rsic indices are
  common on the main sequence, but the reverse is not true---there are
  few low-$n$ transitional/quenched galaxies. Indeed the lower
  threshold for S\'{e}rsic index off the nain sequence
  ($n \approx 2.5$) coincides with the demarcation between the
  galaxies containing the central spheroid (classical bulge) and the
  ones that do not \citep{drory07}. This confirms previous results
  regarding the structure of quenching or quenched galaxies
  \citep{bell08,mosleh17} and that the compaction starts on the main
  sequence \citep{cheung12,barro17}.

\item The low-redshift sample shows that the galaxies above the main
  sequence are on average more compact (1.25 times higher $n$) than on
  the main sequence. A similar trend has been reported in
  \citet{schiminovich07,wuyts11}, and is consistent with a compaction
  proceeding via a starburst stage
  \citep{schiminovich07,tacchella16,lapiner23}.

\item The transitional region has a similar 68 percentile range of S\'{e}rsic
  indices as the the main sequence or the quiescent region, i.e., it
  is inconsistent with being the mix of the tails of two populations, as such mixing
  would widen the distribution in the transitional region.

\end{enumerate}

\begin{figure}[tb]
\centering
\includegraphics[width=0.9\linewidth, clip]{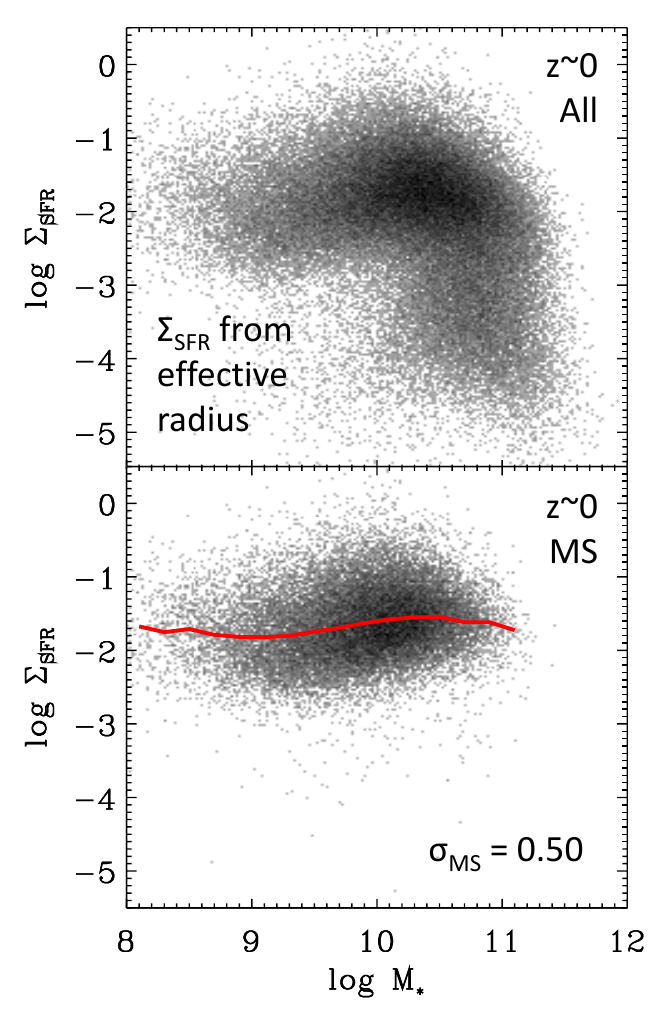}
\caption{The effects on the \sigsfr---$M_*$ relation arising from
  defining the SFR surface density using the effective (half-light)
  radius. This figure is to be compared to the right panel of Figure
  \ref{fig:ssig_sfr_mass}, which uses the isophotal sizes, our nominal
  choice. Effective radius-based \sigsfr\ are significantly noisier,
  leading to a broader main sequence (width, in dex, given in lower
  panel). The effective radius, being dependent on the galaxy light
  profile, is not as good of an indicator of the extent over which SF
  is (or was) taking place as is the isophotal radius.}
\label{fig:sigsfr_reff_mass}
\end{figure}

\section{Discussion}\label{sec:disc}

\subsection{Relevant measure of galaxy size for
  \sigsfr}\label{ssec:reff}

It is common in the literature to obtain the global SFR surface
density either using the effective (half-light) radius, or using the
isophotal radius. We have found that the choice of measure is very
important. In Figure \ref{fig:sigsfr_reff_mass} we show the
\sigsfr---$M_*$ diagram for our low-redshift samples in which \sigsfr\
is based on the effective radius (Eq.\ \ref{eq:area_eff}). This figure
is to be compared to the right panel of Figure
\ref{fig:ssig_sfr_mass}, which was based on the isophotal radius (Eq.\
\ref{eq:area_iso}).  As expected, the absolute values of \sigsfr\ are
higher, because the isophotal sizes are about three times larger on
average than the effective one (Figure
\ref{fig:riso_reff_mass_rpred}). Nonetheless, the relatively flat
shape (and even the details of the inflections) of the \sigsfr\ main
sequence remain.

\begin{figure*}[tb]
\centering
\includegraphics[width=\linewidth, clip]{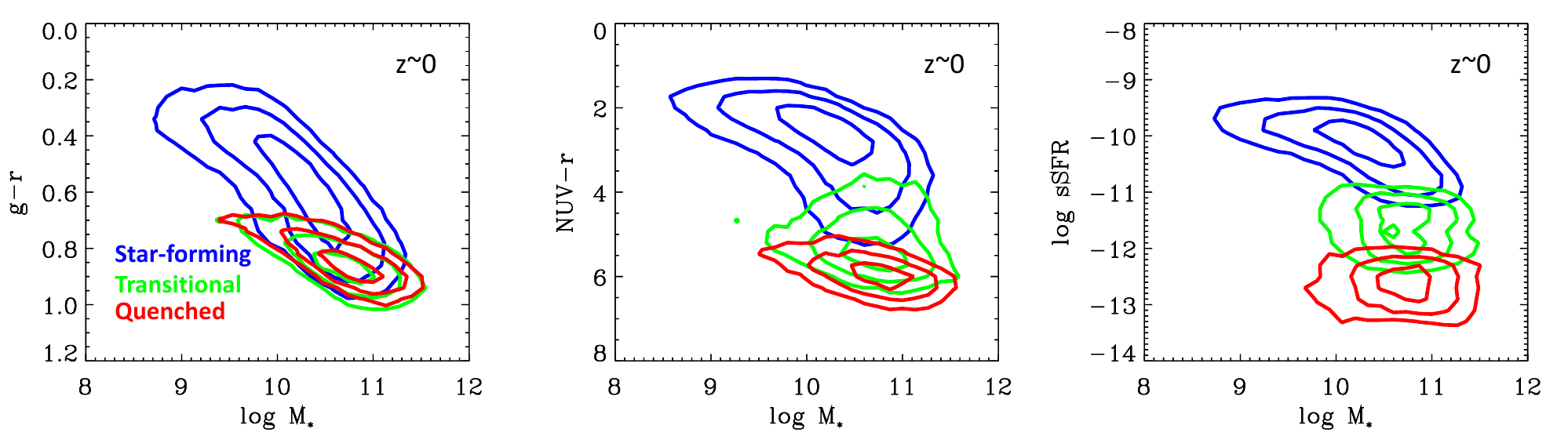}
\caption{The placement of galaxies separated into star-forming,
  transitional and quenched (quiescent) categories on various diagrams
  with stellar mass on the $x$ axis. Panels from left to right
  demonstrate the successive improvements in the ability to separate
  these categories afforded by moving from optical to UV-optical color
  and then to sSFR. Specific SFRs provide a good separation between
  these categories, but the use of \sigsfr\ (based on which these
  categories have been defined) goes an additional step by eliminating
  the effect of age on the main sequence (its tilt).}
\label{fig:vs_mass_contour}
\end{figure*}

What has drastically changed is the the width of the \sigsfr\ main
sequence. Its scatter around the mean is 0.50 dex, compared to 0.28
dex for the main sequence based on the isophotal radius.  Furthermore,
\sigsfr\ based on the effective radius is less well able to
distinguish between the early and late type galaxies. Overall, it is a
much noisier measure of the level of SF activity than the \sigsfr\
based on the isophotal radius. This worse performance cannot be due to
the differences in the precision of the measurements of two sizes. If
anything, the measurements of the effective size are more robust than
the isophotal one \citep{trujillo20,chamba20}. Effective radii used in
this exercise are from the \citet{meert15} catalog. An alternative
source of SDSS effective sizes, the catalog of \citet{simard11},
yields similar results (main sequence scatter of 0.48 dex). Looking at
yet other measures of galaxy size, both \citet{meert15} and
\citet{simard11} catalogs provide estimates for the scale lengths of
the disk components alone. However, \sigsfr\ based on these disk scale
lengths (which are proportional to disk effective radii) produce small
or no improvement in terms of the main sequence scatter compared to
the effective radii of full galaxies.

We propose that the fundamental reason why isophotal sizes produce
more meaningful \sigsfr\ lies in the fact that they better reflect the
extent over which the star formation is taking
place. Isophotal and effective size are actually very different
measures. To first order the isophote reflects a certain mass density
threshold, and is therefore related to the extent over which SF
happens. On the other hand, the effective radius is sensitive to the
galaxy profile, i.e., its structure. This can be illustrated by the
following example. Take two identical bulge-less disk galaxies. Their
isophotal areas are the same. Now let us place in one of them a massive
compact bulge with no ongoing star formation. The effective radius of
that galaxy shrinks, leading to a higher \sigsfr\ if it was based on
the effective size, even though nothing has changed in terms of the
extent over which SF takes place. The isophotal size and \sigsfr\
based on it however, are not affected by this addition of a bulge.

Recently, \citet{trujillo20} have argued against the effective size,
on the grounds that it depends on the light profile, and illustrate
their point with an example similar to one above. Instead, they have
proposed an ``iso-mass'' measure of size based on the mass density
threshold of 1 $M_{\odot} \mathrm{kpc}^{-2}$. Isophotal sizes are a
good proxy for iso-mass sizes, as confirmed by \citet{tang20}, who
found that the optical colors of different galaxies at 25 AB mag
arcsec$^{-2}$ in $g$-band are quite uniform, implying, via
color---$M/L$ relation, that such isophote is a good tracer of stellar
mass surface density. Our nominal isophotal size is based on 25 AB mag
arcsec$^{-2}$ in $r$-band, which is somewhat deeper than the
traditional 25 Vega mag arcsec$^{-2}$ in $B$-band. We find that the
isophotal sizes based on $u$ and $g$-band produce almost identical
results as $r$ in terms of the width of the main sequence and the
ability to distinguish early-type galaxies, even though they
correspond to somewhat smaller (0.22 dex and 0.06 dex, respectively)
physical sizes than the $r$-band isophote. Conceptually, it is not
clear that one should aim to use the area based on the very short
wavelengths. Tying the SFR surface density to an area over which a SF
is or could be taking place (if gas was present), which is achieved by
using optical sizes, allows us to incorporate both the actively
star-forming and the quiescent galaxies (for which SFRs are
essentially upper limits and UV sizes would be meaningless) into a
single scheme.

We conclude that any sort of ``iso'' size (isophotal or iso-mass) will
be more appropriate as a basis for \sigsfr\ as an indicator of SF
activity than an effective size. Indeed, the original global
Kennicutt-Schmidt relation is based on the \sigsfr\ from the isophotal
size \citep{kennicutt89,buat89}, and the rationale for this choice
given in \citet{kennicutt98} was that the isophotal radius is
comparable to the extent of active SF disk in H$\alpha$. For these and
other reasons that suggest that isophotal sizes are better behaved and
provide tighter scaling relations (e.g.,
\citealt{saintonge11,tang20}), future surveys should aim to include
them in their catalogs. If that is not possible (for example due to
the difficulties arising from the cosmological dimming), a viable
alternative would be to estimate the isophotal size from the
combination of the effective size and the stellar mass (Section
\ref{ssec:sigsfr}). Indeed, using this calibration to infer isophotal
sizes essentially recovers the tightness of the \sigsfr\ main sequence
(0.30 dex, vs.\ 0.28 dex with the actual isophotal size).

\subsection{Towards a more physical measure of SF activity}

Being tied to the gas densities and the effectiveness of stellar
feedback, the \sigsfr\ may be considered as a move towards a more physical
measure of the current SF activity/quiescence. In that sense, a switch
to \sigsfr\ aims to provide further conceptual and practical
improvements, similar to one that sSFR had with respect to the use of
optical color. As illustrated in the left panel of Figure
\ref{fig:vs_mass_contour}, the optical color of actively star-forming
galaxies is even more strongly affected by the age of stellar
populations than sSFR, and produces a steep tilt when plotted against
the mass. Furthermore, the optical color has very poor sensitivity to
low relative levels of SF (e.g., \citealt{kauffmann07}), which results
in an inability to distinguish between transitional and fully quenched
galaxies \citep{s14}. As a result, even the massive main-sequence
galaxies have optical colors nearly as red as the early-type galaxies
\citep{cortese12}, affecting the quenched fraction estimates based on
optical color (Figure \ref{fig:vs_mass_contour}, left panel) and
making it appear as if there were no, or very few, massive
star-forming galaxies.  UV-optical color overcome many of the
limitations of the optical colors (middle panel of Figure
\ref{fig:vs_mass_contour}), but is subject to dust and metallicity
effects (which can somewhat be mitigated by combining UV-optical color
with a optical-near-IR color, e.g., in NUV--$r$--$J$
diagram, \citealt{arnouts13}). sSFR determined from the SED fitting
effectively utilizes a range of UV-optical-near-IR color, but being
constrained using the models that include dust and metallicity
effects, is not subject to them. As a result, it provides a cleaner
separation between star-forming, transitional and quenched galaxies
(Figure \ref{fig:vs_mass_contour}). Finally, \sigsfr\ improves over
sSFR by removing the age effect, which through $M_*$ or red optical
band was present in all previous measures discussed here.

\citet{kennicutt12} considered \sigsfr\ as one of the two ways to
normalize SFR, the other being the sSFR. They also commented that the
range of \sigsfr\ for normal (non-starburst) galaxies is relatively
small and in that sense similar to the range of sSFR. More
quantitatively, we see that the observed \sigsfr\ main sequence has
the same scatter as the sSFR main sequence (0.28 dex, Section
\ref{ssec:lowzms}), and intrinsically, the \sigsfr\ main sequence may
be even narrower than the sSFR one. Namely, since we measure the
scatter in small mass bins, the only contributor to sSFR measurement
uncertainty is that of SFR. On the other hand, \sigsfr\ has
measurement uncertainties from both the SFR and the isophotal area.
This is an indication that \sigsfr\ may be a more physical measure of
current SF activity than sSFR.

Furthermore, replacing sSFR with \sigsfr\ almost entirely removes the
downward tilt of the main sequence at low redshift and even results in
a slight upward trend (Figure \ref{fig:ssig_sfr_mass}). The downward
tilt of the main sequence in sSFR--$M_*$ is essentially the result of
SF histories being dependent on the mass, and is unrelated to the
current SF level (see Section \ref{sec:intro}). On the other hand, the
main sequence based on the SFR surface density makes the character of
the ongoing SF more uniform for galaxies of different masses. The
relative constancy of \sigsfr\ across the main sequence was first
pointed out by \citet{schiminovich07}, who called the result
``intriguing''. That result has not received much attention and, as
far as we are aware, was not the focus of any theoretical work. We now
find that at higher redshifts the \sigsfr\ actually rises with the
mass. This is consistent with the rapid growth of the central mass
concentration (a bulge) in more massive galaxies, but less so in
present-day dwarfs and late spirals. Slight upward tilt of the
\sigsfr\ main sequence at $z \sim 0$ may suggest that there is still
some in-situ bulge build-up in massive star-forming
galaxies. We agree with \citet{schiminovich07}, who concluded
that the redshift evolution lies fundamentally in \sigsfr. Indeed, a
galaxy which maintains a constant SFR will be progressively dropping
in sSFR by definition. Considering that \sigsfr\ provides complimentary information
to (s)SFR, we propose that \sigsfr---$M_*$ scaling relation be
included among the benchmarks for galaxy simulations.

A flattening of the main sequence can to some degree be produced by
normalizing SFR not by the total stellar mass, but only the disk
stellar mass, as proposed and shown by \citet{abramson14}. An
underlying assumption behind this modification is that the bulge
represents an inert component that is not associated with the current
star formation. We confirm with our low-redshift sample that replacing
the nominal sSFR with the disk-only sSFR, obtained by multiplying our
total stellar mass with the disk-to-(disk+bulge) mass ratio from the
$n=4+1$ decompositions of \citet{mendel14}, reduces the tilt of the
main sequence from $-0.35$ to $-0.20$ ($\log M_* >8$). The disk sSFR
main sequence is broader than the nominal sSFR one (0.36 dex vs.\ 0.28
dex), most likely because of the greater uncertainties in deriving the
stellar mass of the disk component compared to the total stellar mass.
Disk-bulge decompositions are especially challenging based on SDSS
images. By using the disk mass to normalize SFR, this measure becomes
partially decoupled from the past SF history and therefore has similar
aims as the use of \sigsfr. One conceptual advantage of \sigsfr\ is
that it allows purely spheroidal galaxies with no disk to be
encompassed by the scheme.
 
Many studies nowadays use the relative SFR as a principal variable of
the analysis. Relative SFR will by construction flatten the sSFR main
sequence tilt. It was introduced by \citet{schiminovich07} for the
very reason of eliminating the dependence of sSFR on $M_*$. There is
no ambiguity that in the relative sense (for galaxies of fixed mass)
the galaxies with the high relative SFR can be considered as
experiencing a current burst, whereas the galaxies with low relative
SFR have or are experiencing quenching and have a diminished current
capacity to form stars. Our analysis shows that the relative SFR has a
comparable ability to identify starbursts and early-type galaxies as
does \sigsfr. Its non-optimal aspect is that it is defined relative to
a main sequence that needs to be observationally established, which is
by no means unambiguous, especially at the massive end where the main
sequence blends with the turn off. More importantly, by referring to
SFRs in relative terms, we are in a way giving up on the idea that
there is a physical quantity that describes the SF level.
Interestingly, \citet{schiminovich07} said that the physical basis for
the introduction of the relative SFR comes from its correlation with
\sigsfr.

Global (integrated) SFR surface density is not commonly considered in
the studies of galaxy evolution outside of the context of the
Kennicutt-Schmidt relation. For example, the SF history is usually
defined as the change in SFR over time. \citet{lehnert14}, on the
other hand, discuss the evolution of the MW in terms of the change in
\sigsfr\ (which they call SF intensity, cf.\
\citealt{lanzetta02,boquien10}). They note that it is \sigsfr\ that
determines the role of stellar feedback on outflows and on
mass--metallicity relation.

Likewise, \sigsfr--$M_*$ featuring global SFR density is a rarely used
diagram. \citet{kelly14} and \citet{lunnan15} used it to compare SF
properties of long gamma ray bursts and superluminous supernovae hosts
to that of other supernova hosts (and find them to be elevated.)
\citet{tran17} use it to compare field and cluster galaxies at
$z\sim 2$, and describe \sigsfr\ as the intensity of
SF. \citet{forsterschreiber19} show the \sigsfr--$M_*$ diagram of 600
galaxies at $0.6<z<2.7$ color-coded by incidence of outflows. The
incidence follows \sigsfr\ remarkably well (as pointed out in
\citealt{heckman02,newman12}), and somewhat better than the main
sequence offset (their Fig.\ 7). Interestingly, their \sigsfr--$M_*$
main sequence shows an upward tilt similar to what we saw in the
$z \sim1$ panel of Figure \ref{fig:ssig_sfr_mass_z0_z1_z2}.

\subsection{Implications for studies of resolved star formation}

The advent of the integral field unit (IFU) spectrographs and associated
surveys, such as MaNGA \citep{bundy15}, CALIFA \citep{sanchez16} and
SAMI \citep{bryant15}, has shifted the focus from general considerations
of global SF level to trying to understand the processes of SF
regulation on spatially resolved scales. The most common aspect of IFU
studies concerns the radial profile of SF activity and the question of
the dynamics of the quenching process, such as the inside-out vs.\ the
outside-in scenarios (e.g., \citealt{tacchella15,belfiore18,lin19}).

One can imagine making two types of radial profiles that involve
SFR-related quantities. One is the sSFR radial plot, where the SFR in
a radial bin is divided by the stellar mass in that radial bin
(sometimes designated as $\Sigma_{\mathrm{sSFR}}$), and another is the
\sigsfr\ radial plot, where the SFR is divided by the physical area of
the radial bin. Here we wish to point out that the sSFR radial profile
is subject to the same ambiguities as the global sSFR in the sense
that sSFR depends on both the current SF level and the past SF
history. Consider an example of a galaxy with a prominent bulge, i.e.,
a large stellar mass concentration. A profile of such a galaxy could
likely be redder in the bulge area, corresponding to a dip in the sSFR
profile. However, lower sSFR because of the substantial mass does not
imply that SF levels are suppressed. SF may actually be present in the
bulge at the same or higher levels than further out in the disk
\citep{tacchella18}. There is no question that in such a case the
current SF in the central region contributes relatively little to the
stellar mass compared to when the bulge was being built up, but that
relative change does not imply that any active quenching is taking
place now, rather than just a gradual decline. Thus, for galaxies that
have red (low sSFR) centers but significant amount of SF, the more
neutral term may be the inside-out build-up (e.g.\,
\citealt{nelson16,lilly16,belfiore18}), rather than the inside-out
quenching.

\begin{figure}[tb]
\centering
\includegraphics[width=1.0\linewidth, clip]{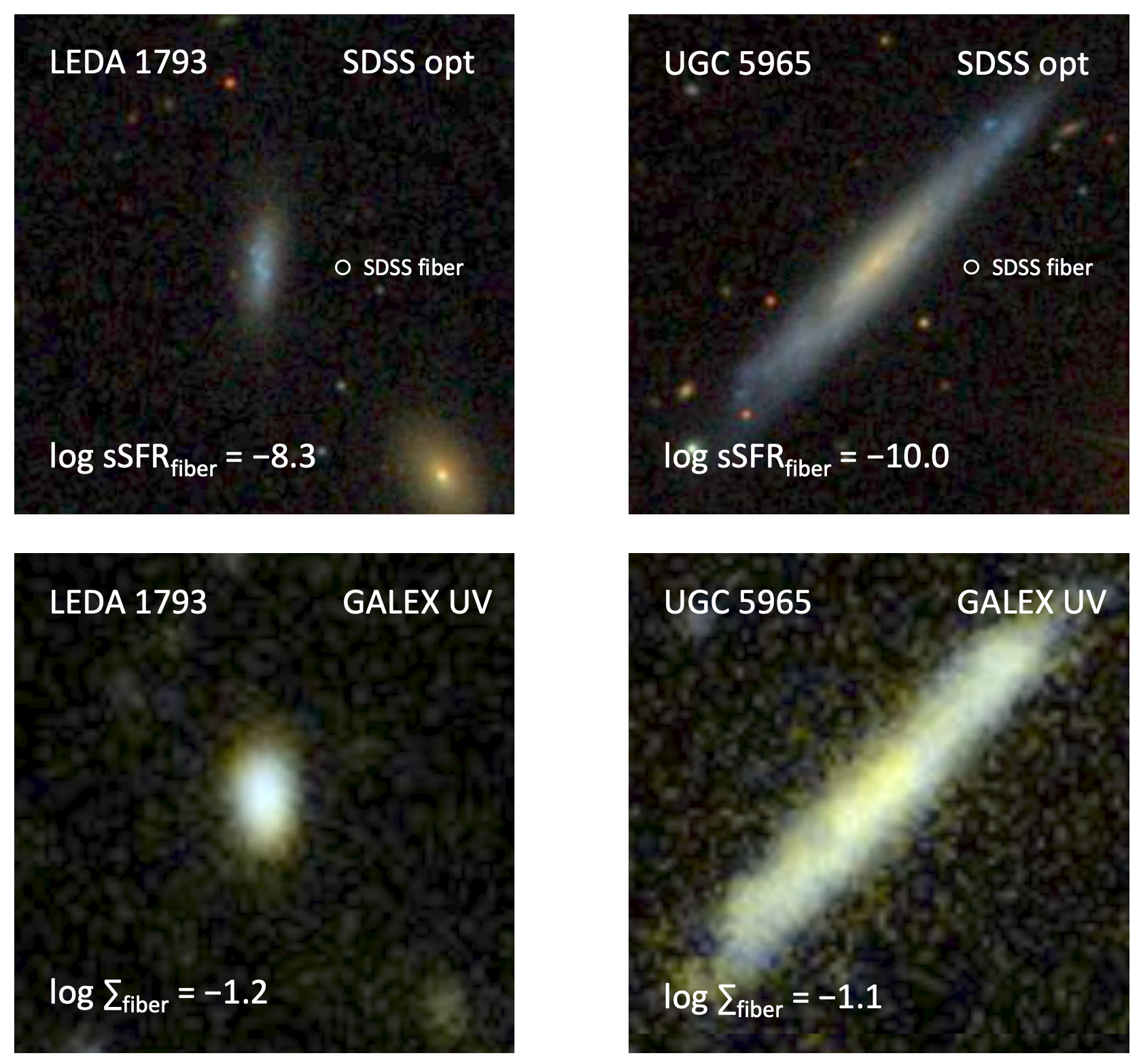}
\caption{Differences in inferences regarding a galaxy's star formation
  based on whether sSFR or \sigsfr\ is considered. Left panels
  correspond to LEDA 1793, whereas right panels show UGC 5965. Upper
  panels present optical images from SDSS and the lower panels are UV
  images from GALEX. UGC 5965 has a red center, but its central SFR
  surface density is as high as that of LEDA 1793, which is blue and
  has a high central sSFR. Quantities are based on dust-corrected
  H$\alpha$ and correspond to 3 arcsec SDSS fiber, shown as a
  circle. SDSS images are taken from SkyServer, whereas GALEX images
  (far-UV plus near-UV composites) are from the Legacy Survey Sky
  Browser.}
\label{fig:images}
\end{figure}

We illustrate this point with a specific example in Figure
\ref{fig:images}, where we take advantage of sSFR and \sigsfr\
determined from dust corrected H$\alpha$ within the SDSS spectroscopic
fiber to probe central quantities. Galaxy on the left (LEDA 1793) has
a very high central sSFR, appearing as a blue compact dwarf, whereas
sSFR is significantly lower for the galaxy on the right (UGC 5965).
UGC 5965 has a distinctly red bulge in SDSS image. The UV images from
GALEX paint a different picture. UGC 5965 reaches the highest UV
brightness in the bulge. As a matter of fact, the UV surface
brightness appears to be similar in UGC 5965 and LEDA 1793, as
corroborated by nearly identical \sigsfr\ in the fiber.

\section{Conclusions}\label{sec:concl}

Replacing (s)SFR with \sigsfr\ in the galaxy ``HR diagram'' has
given us a different perspective regarding the character of SF on and off
the main sequence, and its evolution. The main findings are:

\begin{enumerate}

\itemsep -0.5\parsep

\item The SFR surface density (\sigsfr) is largely insensitive
  to past SF history and thus provides a measure of the current
  global star-forming level of a galaxy tied to its molecular gas
  density. This is contrast to sSFR, a relative measure of
  young-to-old population. 

\item \sigsfr\ provides a cleaner separation of likely starbursts and
  spheroid-dominated (early-type) galaxies than sSFR. Its selection
  power is comparable to that of the SFR offset relative to the main
  sequence.

\item \sigsfr\ of the main-sequence galaxies at low redshift is
  essentially mass-independent. Dwarfs and high-mass spiral (disk)
  galaxies have very similar levels of star-forming activity. This was
  first pointed out in \citet{schiminovich07}.

\item \sigsfr--$M_*$ main sequence at $z \gtrsim 1$ is tilted upwards:
  high-mass galaxies had $\sim 2\times$ higher SF levels than
  lower-mass galaxies, possibly reflecting a rapid build up of a
  central mass concentration (bulge). Such trend is not seen in
  SFR--$M_*$ or sSFR--$M_*$ diagrams, the slope of which does not
  evolve much between $z=0$ and $z=2$. Because of this
  complementarity, we propose that \sigsfr---$M_*$scaling relation be
  included among the benchmarks for galaxy simulations.

\item We confirm that galaxies that fall below the main sequence at
  high redshift are structurally similar to quenched galaxies today
  \citep{wuyts11}. However, the \sigsfr\ values of many such
  high-redshift galaxies are as high as of the
  main-sequence galaxies today. A high-redshift galaxy can drop more
  than 2 dex below the main sequence and still not be fully quenched.

% \item High-redshift transitional galaxies may not necessarily achieve
%   full quenching even by $z=0$, so It may be more accurate to
%   refer to the SF in such galaxies as ``suppressed'', rather than quenched.

\item While the threshold for the onset of quenching is
  redshift-dependent for either \sigsfr\ or sSFR, the former allows
  one to define an absolute threshold for full quiescence that is
  independent of the redshift. We tentatively propose defining full
  quiescence as log \sigsfr\ $<-4.5$ (in units of
  $M_{\odot} \mathrm{yr}^{-1}\mathrm{kpc}^{-2}$) when using $r=25$
  isophotal sizes, or log \sigsfr\ $<-3.5$ when using effective radii.

\item The use of \sigsfr\ radial profiles allows us to distinguish
  between the galaxies where bulges are red (have a central dip in
  sSFR), but where the SF is still proceeding at high levels, from the cases
  where the SF is suppressed in the centers.
  
\item The ability of \sigsfr\ to serve as a precise measure of SF
  activity is severely affected if the area is based on the effective
  (half-light) radius, rather than the isophotal one. This is because
  the isophotal radius, being tied to the physical mass and gas
  density thresholds, defines the extent over which SF takes place,
  whereas the effective radius depends strongly on galaxy light
  profile/concentration.

\item Isophotal radius can be obtained from a combination of the
  effective size and the stellar mass with an error of just 0.05 dex,
  thus facilitating the use of a more precise form of \sigsfr\ in
  cases where isophotal sizes are not available.

% \item Transitional galaxies have a similar spread of S\'{e}rsic
%   indices as the the main sequence or quiescent galaxies, inconsistent
%   with being the mix of the tails of those two populations.
  
\end{enumerate}

The main takeaway message from this study is that the use of sSFR (or
SFR), especially in the context of (s)SFR--$M_*$ diagram, should be
critically assessed depending on the context, and where appropriate be
complemented with the plots involving \sigsfr. 

\acknowledgments The construction of GSWLC is funded through NASA
awards NNX12AE06G and 80NSSC20K0440. We thank the referee for helpful
comments and Patricio Cubillos for sharing his version of ApJ template
on GitHub.

\begin{appendices}

\section{Main sequence tilt in sSFR--$M_*$ diagram} 
\label{app:other_ms}

\begin{figure*}[tb]
\centering
\includegraphics[width=\linewidth, clip]{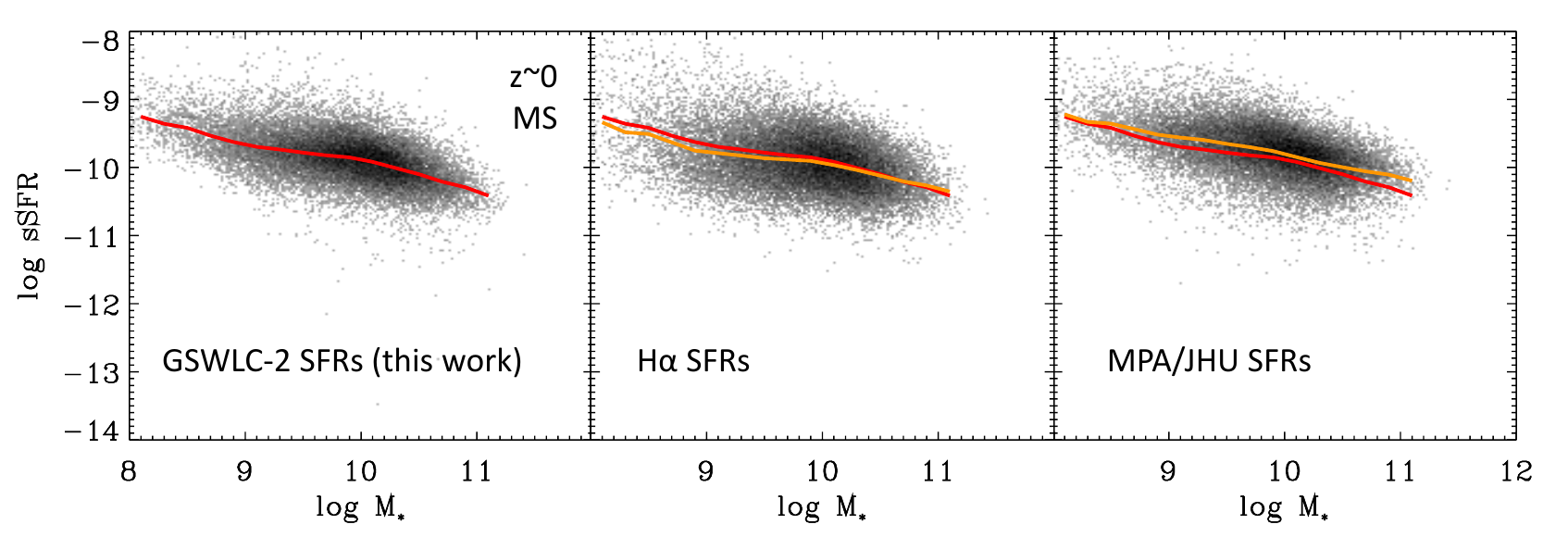}
\caption{Diagrams of the specific SFR vs.\ stellar mass based on three
  measures of the SFR. All panels include the same, main-sequence
  galaxies. Left panel uses integrated sSFRs from the SED fitting
  (nominal values in this work). Running average is shown as a red
  line. Middle panel uses sSFRs from the Balmer decrement-corrected
  H~$\alpha$ emission. Right panel uses SFRs from the MPA/JHU
  catalog. In the middle and right panels the running average is shown
  as an orange line, whereas the running average from the left panel
  is repeated for reference.}
\label{fig:ssfr_mass_other}
\end{figure*}

In Section \ref{ssec:lowzms} we showed that replacing sSFR with
\sigsfr\ has the effect of flattening the main sequence. Here we
discuss how robust is the tilt in the sSFR--$M_*$ diagram in the first
place. In Figure \ref{fig:ssfr_mass_other} we show the
low-redshift main sequence sample using three estimations of
sSFR. Left panel shows our nominal values from GSWLC-2, where SFRs and
$M_*$ are derived from the SED fitting that combines UV and optical
photometry with the constraints on dust from the total IR luminosity
inferred from mid-IR observations and dust emission templates. SFRs
are defined as averages over the last 100 Myr.

The middle panel of Figure \ref{fig:ssfr_mass_other} features SFRs
based on an entirely different SF indicator---the nebular emission
lines. We take H$\alpha$ fluxes and correct them for dust attenuation
using the Balmer decrement and assuming a Galactic extinction
curve. Since the line emission is measured in 3 arcsec spectroscopic
fibers, we use the stellar mass also in the fiber in order to get the
correct sSFR. Despite the fact that the fiber typically includes only
23\% of the galaxy mass, the agreement between these and our
integrated sSFRs is remarkable---even the position of the breaks in
the main sequence matches. There is somewhat more scatter at lower
masses, which may be due to the shorter timescale over which H$\alpha$
SFR is sensitive.

The right panel of Figure \ref{fig:ssfr_mass_other} displays SFRs
from \citet{brinchmann04} as updated in the DR7 version of the MPA/JHU
catalog. These SFRs are the sum of emission-line SFRs in the fiber and
SFRs for the annulus around the fiber aperture, obtained from the
optical SED fitting. For consistency, these SFRs are normalized by the
$M_*$ also from the MPA/JHU catalog. A small offset between the main
sequence from these and our nominal sSFRs is present, but its overall
character and the tilt are similar.

Essentially the same main sequence trend as our nominal one (including
the positon of the breaks) is also obtained from SFRs and stellar
masses from \citet{chang15} catalog (plot not shown), which is based
on the optical/IR SED fitting, but unlike GSWLC-2 includes all four
bands from WISE directly and uses MAGPHYS \citep{dacunha08} models.

\end{appendices}

\end{document}